\documentclass[10pt,english,aps,showkeys,showpacs,nofootinbib]{revtex4-1}
\usepackage[T1]{fontenc}
\usepackage{textcomp}
\usepackage[utf8]{inputenc}
\usepackage{float}
\usepackage{units}
\usepackage{amsmath}
\usepackage{amssymb}
\usepackage{stackrel}
\usepackage{graphicx}

\makeatletter
\ifdefined\showcaptionsetup
 \PassOptionsToPackage{caption=false}{subfig}
\fi
\usepackage{subfig}
\makeatother

\usepackage{babel}
\begin{document}
\selectlanguage{english}
\title{Statistical properties of the two dimensional Feshbach–Villars oscillator
(FVO) in the Rotating Cosmic String space-time}
\author{Abdelmalek Bouzenada}
\email{abdelmalek.bouzenada@univ-tebessa.dz ; abdelmalekbouzenada@gmail.com}

\affiliation{Laboratoire de Physique Appliquée et Théorique~\\
 Université Larbi-Tébessi-, Tébessa, Algeria}
\author{Abdelmalek Boumali}
\email{boumali.abdelmalek@gmail.com}

\affiliation{Laboratoire de Physique Appliquée et Théorique~\\
 Université Larbi-Tébessi-, Tébessa, Algeria}
\date{\today}
\begin{abstract}
This paper is concerned with an investigation of the quantum mechanical
dynamics of massive, spinless relativistic Klein-Gordon particle in
the space-time generated by a rotating cosmic string. The equations
of motion are found by the use of the first-order Feshbach-Villars
formulation of the Klein-Gordon equation. The wave-functions and the
associated energies were deduced ( both in the free and in the interaction
case). Following that, the partition function was approximated using
Zeta function representation. Moreover, by considering the energy
spectrum of the system in question, the thermal properties are presented.
We examine the behavior of these properties as a function of the physical
parameters of the model such as rotation, curvature, and quantum numbers.
Therefore, the impact of the topological defect on the quantum system
under investigation is discussed.
\end{abstract}
\keywords{Klein-Gordon oscillator, Feshbach–Villars equations, topological defects,
cosmic string space-time, partition function, thermal properties.}
\pacs{04.62.+v; 04.40.\textminus b; 04.20.Gz; 04.20.Jb; 04.20.\textminus q;
03.65.Pm; 03.50.\textminus z; 03.65.Ge; 03.65.\textminus w; 05.70.Ce}
\maketitle

\section{Introduction }

It is of general interest to understand the influence of the gravitational
field on the dynamics of quantum mechanical systems. Einstein’s theory
of general relativity (GR) \Citep{key-1}, on the one hand, provides
a persuasive description of gravity as a geometric property of space-time.
In particular, it demonstrates that the classical gravitational field
is a manifestation of the curvature of space-time. It has, for example,
successfully predicted the existence of gravitational waves \Citep{key-2}
and black holes\Citep{key-3}. Quantum mechanics (QM), on the other
hand, is the framework for describing the behavior of particles in
the microscopic scale \Citep{key-4}. It is an extremely successful
theory (typically quantum field theory) in explaining how the tiny
particles interact and how three of the four fundamental forces of
nature: the weak, strong, and electromagnetic interactions are emerged
\Citep{key-5}. However, the attempts to have a unified theory which
can reconcile general relativity and quantum mechanics i.e. a theory
of quantum gravity, is still suffering from several obstacles and
technical issues that cannot being encountered, at least until the
recent time \Citep{key-6,key-7} .

One basic method for formulating a theory in which the interface between
gravity and relativistic quantum mechanics arises, is by generalizing
the aspects of the relativistic dynamics of particles in flat Minkowski
space to an arbitrary curved background geometry \Citep{key-8,key-9}
, and therefore one can extend this formulation to establish a broad
picture on how the gravitational field affects on the relativistic
particles at quantum level. In this way, it is possible to adapt the
approach to deal with different models where the notion of curvature
appears, and hence incorporating more predictions on the values of
macroscopic observables which are required to make relevant experimental
verification of certain phenomenological consequences, specifically
in astrophysics and cosmology. In addition, understanding the thermodynamic
behavior of relativistic particles where gravitational effects have
to be taken into account \Citep{key-10,key-11,key-12}, and analyzing
the associated features i.e, the fundamental statistical quantities,
would offer the possibility of obtaining useful and essential results
in the context of describing the quantum behavior of gravity.

Over the past few decades, topological defects ( domain walls, cosmic
strings, monopoles, and textures ) have been tackled intensively as
a subject of research and still remain one of the most active fields
in condensed matter physics, cosmology, astrophysics and elementary
particle models. It is believed that these structures raised as a
consequence of the Kibble mechanism \citep{key-13,key-14,key-15}
where the defects are formed in symmetry-breaking phase transitions
during the cooling of the early universe \Citep{key-16,key-17}. The
particular defect we are interested in, is cosmic strings (for more
reviews we refer the reader to \Citep*{key-18}). These objects (
static or rotating) can produce observable effects. For example, they
provide a promising scheme for seeding galaxy formation and the gravitational
lensing effects. Moreover, by studying cosmic strings and their properties,
we can learn much more about particle physics at very high energies
in different scenarios. In addition, the possibility that cosmic strings
could behave like superconducting wires has been raised with intriguing
consequences in modern physics.

It has been known for many years now that the harmonic oscillator
(HO) is considered as an essential tool in many disciplines of theoretical
physics \Citep{key-19}. It is a well studied exactly solvable model
which can be used to analyze various complex problems in the framework
of quantum mechanics \Citep{key-20}. Furthermore the relativistic
generalization of the quantum harmonic oscillator yields an effective
model in explaining very diverse aspects of molecular, atomic and
nuclear interactions. Indeed, the feature of having a complete set
of exact analytical solutions when dealing with such model can give
rise to significantly different explanations of many mathematical
and physical phenomena, and hence related applications can be achieved
via the underlying formulation. 

It is now accepted that the behavior of several relativistic quantum
systems depends crucially on the so-called Dirac oscillator (DO).
As pointed out by It\^{o} et al. \citep{key-21} in earlier developments
of spin-1/2 particle dynamics with linear trajectory. They showed
that the non-relativistic limit of this system leads to the ordinary
harmonic oscillator with a strong spin-orbit coupling term. Actually,
according to Moshinsky and Szczepaniak \citep{key-22} the above mentioned
DO could be obtained from the free Dirac equation by introducing an
external linear potential through a minimal substitution of the momentum
operator $\hat{p}\longrightarrow\hat{p}-im\omega\beta\hat{r}$. It
is interesting to point out that besides the theoretical focus on
studying the DO, valuable insights can be obtained by considering
the physical interpretation which is certainly essential in understanding
many relevant applications.

Inspired by DO, an analogous formalism was introduced for the case
of bosonic particles and hence it was named a Klein-Gordon oscillator
(KGO) \citep{key-23,key-24}. The covariant form of this model in
curved space-time and within different configurations, has recently
been an active field of investigation by several authors. There are
numerous contributions on the subject of the relativistic quantum
motions of scalar and vector particles under gravitational effects
produced by different curved space-time geometries, for example, the
problem of the interaction between KGO coupled harmonically with topological
defects in Kaluza-Klein theory is studied in Ref. \Citep{key-25}.
The relativistic quantum dynamics of spin-0 particles in a rotating
cosmic string space-time with a scalar and vector potentials of Coulomb-type
has been studied in Ref. \Citep{key-26}. Furthermore, rotating effects
on the scalar field in the cosmic string space-time, in the space-time
with space-like dislocation and in the space-time with a spiral dislocation
have been investigated in Ref. \Citep{key-27}. Recently, the authors
of Ref. \Citep{key-28} have analyzed the KGO in a cosmic string space-time
and studied the effects stemming from the rotating frame and non-commutativity
in momentum space. In addition, the KGO subjected to a magnetic quantum
flux in the presence of a Cornell-type scalar and Coulomb-type vector
potentials in a rotating cosmic string space-time have been examined
in Ref. \Citep{key-29}. 

Attempts to investigate the relativistic spin-0, spin-1 bosons and
spin-1/2 fermions wave functions and their time evolution have been
pursued by various authors \Citep{key-30,key-31,key-32} making use
of the Hamiltonian form i.e, having Schrodinger's equations. The so-called
Feshbach-Villars (FV) equations \Citep{key-33} are of particular
interest in this respect. These equations were initially constructed
in the purpose of permitting a relativistic single particle interpretation
of the second order KG equation. For the later case, FV equations
originate from splitting the KG wave function into two components
in order to obtain an equation with first order time derivative. In
recent decades, a number of papers have been produced with the aim
of exploring the relativistic dynamical properties of single particles
and solving their wave equations by adopting the FV scheme (e.g, Refs.
\Citep{key-34,key-35,key-36,key-37,key-38,key-39,key-40} and other
related references cited therein). 

In a quest to map out a description of the dynamics of spinless, massive
bosons in a space-time with a topological defect, namely the cosmic
string geometry, we employ the methods of the FV transformation to
derive FV equations of motion in a static and rotating cosmic string
backgrounds. The main purpose of this paper is , firstly to solve
the derived FV equations and obtain the associated spectra, and secondly,
to investigate the thermodynamic properties of these systems and understand
how the geometry of the space-time affects on the observable quantities.

The structure of this paper is as follows. In the next section we
derive the FV equations for scalar boson in Minkowski and static Cosmic
string space-time considering both the free and the interaction case.
We introduce the KG oscillator in a Hamiltonian form, then we solve
the obtained equations to deduce the eigenstates and the energy levels.
In section 3 we furthermore construct the previous equations in the
case where the geometry of the cosmic string is characterized by a
rotation, the same procedure of Section 2 is used in this section.
In section 4 we study the statistical properties of the system in
question and we discuss the influence of the topological defect on
these quantities. We give our conclusions in section 5. Throughout
the paper, we will always use natural units $\hbar=c=G=K_{B}=1$ ,
and our metric convention is $(+,-,-,-$).

\section{The FV Representation of Spin-0 Particle in Minkowski Space-time\protect\label{sec:2}}

\subsection{Klein-Gordon Equation in Schrödinger Form}

In this section, we deal with the relativistic quantum description
of spin-0 particle propagating in Minkowski space-time with the metric
tensor $\eta_{\mu\nu}=\text{diag}\left(1,-1,-1,-1\right)$. For a
scalar massive particle $\Phi$ with mass $m>0$, the standard covariant
KG equation reads \Citep{key-41,key-42}
\begin{equation}
\left(\eta^{\mu\nu}D_{\mu}D_{\nu}+m^{2}\right)\Phi(x,t)=0,\label{eq:1}
\end{equation}
where $D_{\mu}=i\left(p_{\mu}-eA_{\mu}\right)$ denotes the minimally
coupled covariant derivative. $p_{\mu}=\left(E,-p_{i}\right)$ is
the canonical four momentum and $A_{\mu}=\left(A_{0},-A_{i}\right)$
is the electromagnetic four potential, respectively. $e$ is the magnitude
of the particle charge.

At this place, it is worth emphasizing that Eq. (\eqref{eq:1}) can
be rewritten in a Hamiltonian form with the time first derivative
i.e, a Schrodinger's equation
\begin{equation}
\mathcal{H}\Phi\left(x,t\right)=i\,\frac{\partial}{\partial t}\Phi(x,t).\label{eq:2}
\end{equation}
The Hamiltonian $\mathcal{H}$ can be defined by the use of the FV
linearization procedure i.e, transforming Eq. (\ref{eq:1}) to a first
order in time differential equation. We introduce the two component
wave function \Citep{key-43,key-44}, 
\begin{equation}
\Phi(x,t)=\left(\begin{array}{c}
\phi_{1}(x,t)\\
\phi_{2}(x,t)
\end{array}\right)=\frac{1}{\sqrt{2}}\left(\begin{array}{c}
1+\frac{i}{m}\mathcal{D}\\
1-\frac{i}{m}\mathcal{D}
\end{array}\right)\psi(x,t),\label{eq:3}
\end{equation}
here $\psi(x,t)$ obeys the KG wave equation, and we have defined
$\mathcal{D}$ such that
\begin{equation}
\mathcal{D}=\frac{\partial}{\partial t}+ieA_{0}(x),
\end{equation}
the above transformation (\eqref{eq:3}) consists in introducing wave
functions satisfying the conditions 
\begin{equation}
\psi=\phi_{1}+\phi_{2},\qquad i\mathcal{D}\psi=m\left(\phi_{1}-\phi_{2}\right).
\end{equation}
For our subsequent review, it is convenient to write
\begin{equation}
\begin{aligned}\phi_{1} & =\frac{1}{2m}\left[m+i\frac{\partial}{\partial t}-eA_{0}\right]\psi\\
\phi_{2} & =\frac{1}{2m}\left[m-i\frac{\partial}{\partial t}+eA_{0}\right]\psi,
\end{aligned}
\end{equation}
Equivalently, Eq. (\eqref{eq:1}) becomes 
\begin{equation}
\begin{aligned}\left[i\frac{\partial}{\partial t}-eA_{0}\right]\left(\phi_{1}+\phi_{2}\right) & =m\left(\phi_{1}-\phi_{2}\right)\\
\left[i\frac{\partial}{\partial t}-eA_{0}\right]\left(\phi_{1}-\phi_{2}\right) & =\left[\frac{\left(p_{i}-eA_{i}\right)^{2}}{m}+m\right]\left(\phi_{1}+\phi_{2}\right),
\end{aligned}
\end{equation}
Addition and subtraction of these two equations lead to the system
of coupled differential equations of first order in time,
\begin{equation}
\begin{aligned}\frac{\left(p_{i}-eA_{i}\right)^{2}}{2m}\left(\phi_{1}+\phi_{2}\right)+\left(m+eA_{0}\right)\phi_{1} & =i\frac{\partial\phi_{1}}{\partial t}\\
\frac{-\left(p_{i}-eA_{i}\right)^{2}}{2m}\left(\phi_{1}+\phi_{2}\right)-\left(m-eA_{0}\right)\phi_{2} & =i\frac{\partial\phi_{2}}{\partial t},
\end{aligned}
\label{eq:8}
\end{equation}
Using Eqs. (\eqref{eq:8}), the FV Hamiltonian of a scalar particle
in the presence of the electromagnetic interaction may be written
as
\begin{equation}
\mathcal{H}_{KG}=\left(\tau_{z}+i\tau_{y}\right)\frac{\left(p_{i}-eA_{i}\right)^{2}}{2m}+m\tau_{z}+eA_{0}(x),\label{eq:9-1}
\end{equation}
where $\tau_{i}\,\left(i=x,y,z\right)$ are the conventional $2\times2$
Pauli matrices given by
\begin{equation}
\tau_{x}=\left(\begin{array}{cc}
0 & 1\\
1 & 0
\end{array}\right),\quad\tau_{y}=\left(\begin{array}{cc}
0 & -i\\
i & 0
\end{array}\right),\quad\tau_{z}=\left(\begin{array}{cc}
1 & 0\\
0 & -1
\end{array}\right).
\end{equation}
It is noteworthy that the Hamiltonian (\eqref{eq:9-1}) fulfills the
generalized hermicity condition \footnote{The Hamiltonian $\mathcal{H}$ is said to be pseudo-Hermitian if there
is an invertible, Hermitian, linear operator $\beta$ such that $\mathcal{H}^{\dagger}=\beta\mathcal{H}\beta^{-1}$
\Citep{key-45}},
\begin{equation}
\mathcal{H}_{KG}=\tau_{z}\mathcal{H}_{KG}^{\dagger}\tau_{z},\qquad\mathcal{H}_{KG}^{\dagger}=\tau_{z}\mathcal{H}_{KG}\tau_{z}.
\end{equation}
For the free particle propagation i.e, no interaction is considered
$\left(A_{\mu}=0\right),$ the one dimensional FV Hamiltonian simplifies
to
\begin{equation}
\mathcal{H}_{0}=\left(\tau_{z}+i\tau_{y}\right)\frac{p_{x}^{2}}{2m}+m\tau_{z},\label{eq:12-1}
\end{equation}
The solutions to the free Hamiltonian (which is independent of time)
are just stationary states. Assuming a solution of the form \citep{key-36},
\begin{equation}
\Phi\left(x,t\right)=\Phi\left(x\right)e^{-iEt}=\left(\begin{array}{c}
\phi_{1}\left(x\right)\\
\phi_{2}\left(x\right)
\end{array}\right)e^{-iEt},\label{eq:13-1}
\end{equation}
with $E$ being the energy of the system. Thus, Eq. (\eqref{eq:2})
can then be written as
\begin{equation}
\mathcal{H}_{0}\Phi\left(x\right)=E\Phi\left(x\right),\label{eq:14-1}
\end{equation}
which is the one-dimensional FV equation of the free relativistic
spin-0 particle and it is carried out with the aim of having an alternative
Schrodinger's to KG equation. In what follows, the above method will
be used to find the solutions of wave equations in curved space-time,
namely, the cosmic string.

\subsection{KGO in Schrödinger Form}

For the next discussion, it is preferable to analyze the KGO in Minkowski
space-time exploiting the method described previously. In order to
examine the KGO in the FV representation, we start with the minimal
substitution $p_{\mu}\longrightarrow p_{\mu}+im\omega x_{\mu}$ of
the momentum operator \Citep{key-46}. Here $\omega$ is the oscillation
frequency and $x_{\mu}=(0,x,0,0)$ . Thus, by generalizing the momentum
operator, the one dimensional FV Hamiltonian (\eqref{eq:12-1}) becomes
\begin{equation}
\mathcal{H}_{KGO}=\frac{1}{2m}\left(\tau_{z}+i\tau_{y}\right)\left(P_{x}-im\omega x\right)\left(P_{x}+im\omega x\right)+m\tau_{z},\label{eq:15-1}
\end{equation}
In this way, one obtains the Schrödinger formulation of KGO in one
dimension by substituting the Ansatz (\eqref{eq:13-1}) into Eq. (\eqref{eq:14-1})
by means of Eq. (\eqref{eq:15-1}),
\begin{equation}
\begin{aligned}\frac{1}{2m}\left[\frac{d^{2}}{dx^{2}}-m^{2}\omega^{2}x^{2}+m\omega\right]\left(\phi_{1}+\phi_{2}\right)+m\phi_{1} & =E\phi_{1}\\
-\frac{1}{2m}\left[\frac{d^{2}}{dx^{2}}-m^{2}\omega^{2}x^{2}+m\omega\right]\left(\phi_{1}+\phi_{2}\right)-m\phi_{2} & =E\phi_{2},
\end{aligned}
\end{equation}
after performing a calculation similar to the one was done in the
previous KG equation case, then it follows that
\begin{equation}
\left[\frac{d^{2}}{dx^{2}}-m^{2}\omega^{2}x^{2}+\left(E^{2}-m^{2}+m\omega\right)\right]\psi\left(x\right)=0,\label{eq:17-1}
\end{equation}
or,
\begin{equation}
\left[\frac{d^{2}}{dx^{2}}-\lambda_{1}x^{2}+\lambda_{2}\right]\psi\left(x\right)=0\label{eq:18}
\end{equation}
where we have set,
\begin{equation}
\lambda_{1}=m^{2}\omega^{2},\qquad\lambda_{2}=E^{2}-m^{2}+m\omega.
\end{equation}
Eq. (\eqref{eq:18}) is a second order differential equation for the
field $\psi$ describing the KGO dynamics in one dimensional Minkowski
space-time. The solution of the above equation can be found in literature,
and a quantization condition of the energy is followed from that solution,
providing
\begin{equation}
\frac{E^{2}-m^{2}+m\omega}{2m}=\left(n+\frac{1}{2}\right)\omega,\qquad n=0,1,2,\cdots.\label{eq:16}
\end{equation}
After arranging and simplifying the condition (\eqref{eq:16}), we
find the following expression of the energy spectrum
\begin{equation}
E=\pm\sqrt{2m\omega n+m^{2}},\label{eq:17}
\end{equation}
Eq.(\eqref{eq:17}) presents the relativistic energy spectrum of KGO
in Minkowski space-time.

\section{The FV Representation of Spin-0 Particle in Cosmic String Space-time\protect\label{sec:3}}

The purpose of this section is to study the KGO in the background
geometry of a cosmic string with the use of the FV scheme. It is well
known that the generally covariant relativistic wave equations of
a scalar particle in a Riemannian space-time defined by the metric
tensor $g_{\mu\nu}$ , can be obtained by reformulating the KG equation
such that ( see, e.g, the textbooks \Citep{key-8,key-9} )
\begin{equation}
\left(\square+m^{2}-\xi R\right)\Phi(x,t)=0,\label{eq:22-1}
\end{equation}
where $\square$ is the Laplace-Beltrami operator given by
\begin{equation}
\square=g^{\mu\nu}D_{\mu}D_{\nu}=\frac{1}{\sqrt{-g}}\partial_{\mu}\left(\sqrt{-g}g^{\mu\nu}\partial_{\nu}\right),
\end{equation}
$\xi$ is a real dimensionless coupling constant and $R$ is the Ricci
scalar curvature defined by $R=g^{\mu\nu}R_{\mu\nu}$ where $R_{\mu\nu}$
is the Ricci curvature tensor. $g^{\mu\nu}$ is the inverse metric
tensor and $g=\det\left(g^{\mu\nu}\right)$.

We now would like to study the quantum dynamics of spin-0 particles
in the space-time induced by a (2+1)-dimensional static cosmic string
and set up the corresponding FV formulation. 

\subsection{FV formulation of KGO in static cosmic string space-time}

Before we study the KGO in the Hamiltonian representation, let us
first derive the KG wave equation for the free relativistic scalar
particle propagating in the cosmic string space-time that is assumed
to be static and cylindrical symmetric.

The general expression for a (3+1)-dimensional cosmic string metric
is defined by the line element \citep{key-47,key-48} 
\begin{align}
ds^{2} & =g_{\mu\nu}dx^{\mu}dx^{\nu}\nonumber \\
 & =dt^{2}-dr^{2}-\alpha^{2}r^{2}d\varphi^{2}-dz,\label{eq:24}
\end{align}
in cylindrical coordinates \footnote{Note that this metric is an exact solution to Einstein's field equations
for $0\le\mu<1/4,$ and by setting $\varphi^{\prime}=\alpha\varphi$,
then it represents a flat conical exterior space with angle deficit
$\delta\phi=8\pi\mu$.}. Here $-\infty\le t\le+\infty$, $r\ge0$, $0\le\varphi\le2\pi$,
$-\infty\le z\le+\infty,$ and $\alpha\in[0,1[\:$ is the angular
parameter which determines the angular deficit $\delta\varphi=2\pi(1-\alpha)$,
and it is related to the linear mass density $\mu$ of the string
by $\alpha=1-4\mu$. 

For the sake of simplicity in treating our quantum mechanical problem,
let us work in a lower dimensional space in which $z=\text{cont}$,
and because there is no structure in $z$-direction, we can suppress
it \citep{key-49,key-50}. Thus, the metric of a static cosmic string
with cylindrical symmetry has a $(2+1)$-dimensional form \footnote{Since this metric is Lorentz-invariant under boosts in the $(t,z)$
plane \Citep{key-18,key-48}, and by the virtue of rotational symmetry
along the $z$-axis, it is reasonable to assume that the theory is
invariant in the 2-dimentional $(r,\varphi)$ surface.}
\begin{equation}
ds^{2}=dt^{2}-dr^{2}-\alpha^{2}r^{2}d\varphi^{2}\label{eq:25}
\end{equation}
where the components of the metric and the inverse metric tensors
are, respectively,
\begin{equation}
g_{\mu\nu}=\left(\begin{array}{ccc}
1 & 0 & 0\\
0 & -1 & 0\\
0 & 0 & -\left(\alpha r\right)^{2}
\end{array}\right),\quad g^{\mu\nu}=\left(\begin{array}{ccc}
1 & 0 & 0\\
0 & -1 & 0\\
0 & 0 & \frac{-1}{\left(\alpha r\right)^{2}}
\end{array}\right)\label{eq:19}
\end{equation}
Its is worthy to mention that the subject of spinless massive particles
in the geometry generated by a static cosmic string background has
been discussed in several papers ( see, e.g, \Citep{key-25,key-51})

In what follows, we shall adopt the procedure presented in Refs. \Citep{key-52,key-53}
to derive the FV form of KG wave equation in curved manifolds. We
use the generalized Feshbach–Villars transformation (GFVT) \footnote{An equivalent transformation was proposed earlier in Ref.\Citep{key-54}}
which is appropriate for describing both massive and massless particles.
In the GFVT, the components of the wave function $\Phi$ are given
by \citep{key-52} 
\begin{equation}
\psi=\phi_{1}+\phi_{2},\qquad i\tilde{\mathcal{D}}\psi=\mathcal{N}\left(\phi_{1}-\phi_{2}\right),
\end{equation}
where $\mathcal{N}$ is an arbitrary nonzero real parameter, and we
have defined $\tilde{\mathcal{D}}=\frac{\partial}{\partial t}+\mathcal{Y},$
with 
\begin{equation}
\mathcal{Y}=\frac{1}{2g^{00}\sqrt{-g}}\left\{ \partial_{i},\sqrt{-g}g^{0i}\right\} ,\label{eq:25-1}
\end{equation}
The curly bracket in Eq. (\eqref{eq:25-1}) denotes the anti-commutator.
For the above mentioned transformation, the Hamiltonian reads 
\begin{equation}
\mathcal{H}_{GFVT}=\tau_{z}\left(\frac{\mathcal{N}^{2}+\mathcal{T}}{2\mathcal{N}}\right)+i\tau_{y}\left(\frac{-\mathcal{N}^{2}+\mathcal{T}}{2\mathcal{N}}\right)-i\mathcal{Y},\label{eq:20}
\end{equation}
with
\begin{equation}
\mathcal{T}=\frac{1}{g^{00}\sqrt{-g}}\partial_{i}\sqrt{-g}g^{ij}\partial_{j}+\frac{m^{2}-\xi R}{g^{00}}-\mathcal{Y}^{2},\label{eq:30}
\end{equation}
here and below $(i,j=1,2,3)$. We note that for $\mathcal{N}=m$,
the original FV transformations are satisfied.

Now, considering the metric (\ref{eq:25}), it is easy to find that
$R=0$, in other words, the space-time is locally flat (there is no
local gravity), and hence the coupling term is vanishing\footnote{The case $\xi=0$ is refereed to as minimal coupling. However, for
massless theory, $\xi$ takes the value 1/6 (in 4 dimensions). Then,
in this later case, the equations of motion are conformally invariant.}.

A straightforward calculation leads to $\mathcal{Y}=0,$ then we obtain
\begin{equation}
\mathcal{T}=-\frac{d^{2}}{dr^{2}}-\frac{1}{r}\frac{d}{dr}-\frac{1}{\alpha^{2}r^{2}}\frac{d^{2}}{d\varphi^{2}}+m^{2},\label{eq:21}
\end{equation}
Using these results to find the Hamiltonian (\ref{eq:20}), then because
of the time and angular independence in the metric (\ref{eq:25}),
one can assume a solution of the form \footnote{we seek solutions that are cylindrical symmetric, i.e. solutions that
have a rotational symmetry in the $(x,y)$-plane and do not explicitely
depend on $z$.},
\begin{equation}
\Phi(t,r,\varphi)=\Phi(r)e^{-i\left(Et-j\varphi\right)},
\end{equation}
where $j=0,\pm1,\pm2,..$ are the eigenvalues of the $z$component
of the angular momentum operator. Therefore, it follows that the KG
equation (\ref{eq:22-1}) may be written equivalently to the following
two coupled equations
\begin{align}
\left(\mathcal{N}^{2}+\mathcal{T}\right)\phi_{1}+\left(-\mathcal{N}^{2}+\mathcal{T}\right)\phi_{2} & =2\mathcal{N}E\phi_{1}\nonumber \\
-\left(\mathcal{N}^{2}+\mathcal{T}\right)\phi_{2}-\left(-\mathcal{N}^{2}+\mathcal{T}\right)\phi_{1} & =2\mathcal{N}E\phi_{2},\label{eq:22}
\end{align}
The sum and the difference of the two last equations give a second
order differential equation for the field $\psi$. Thus, the radial
equation is written as follows
\begin{equation}
\left[\frac{d^{2}}{dr^{2}}+\frac{1}{r}\frac{d}{dr}-\frac{\zeta^{2}}{r^{2}}+\kappa\right]\psi\left(r\right)=0,\label{eq:23}
\end{equation}
where we have set
\begin{equation}
\zeta=\frac{j}{\alpha},\qquad\kappa=\sqrt{E^{2}-m^{2}}.
\end{equation}
We can observe that Eq. (\eqref{eq:23}) is a Bessel equation and
its general solution is defined by \citep{key-55}
\begin{equation}
\psi\left(r\right)=A\,J_{|\zeta|}\left(\kappa r\right)+B\,Y_{|\zeta|}\left(\kappa r\right),\label{eq:26}
\end{equation}
where $J_{|\zeta|}\left(\kappa r\right)$ and $Y_{|\zeta|}\left(\kappa r\right)$
are the Bessel functions of order $\zeta$ and of the first and the
second kind, respectively. Here $A$ and $B$ are arbitrary constants.
We notice that at the origin when $\zeta=0$, the function $J_{|\zeta|}\left(\kappa r\right)\ne0$.
However, $Y_{|\zeta|}\left(\kappa r\right)$ is always divergent at
the origin. In this case, we will consider only $J_{|\zeta|}\left(\kappa r\right)$
when $\zeta\ne0$. Hence, we write the solution to Eq. (\eqref{eq:23})
as follows
\begin{equation}
\psi\left(r\right)=A\,J_{\frac{|j|}{\alpha}}\left(\sqrt{E^{2}-m^{2}}\,r\right),
\end{equation}
using this solution, we can now write the complete two-components
wave-function of the spinless massive KG particle in the space-time
of a static cosmic string 
\begin{equation}
\psi\left(t,r,\varphi\right)=|C|\left(\begin{array}{c}
1+\frac{E}{\mathcal{N}}\\
1-\frac{E}{\mathcal{N}}
\end{array}\right)e^{-i\left(Et-j\varphi\right)}\,J_{\frac{|j|}{\alpha}}\left(\sqrt{E^{2}-m^{2}}\,r\right),\label{eq:28}
\end{equation}
The constant $|C|$ can be obtained by the appropriate normalization
condition associated with the KG equation ( e.g, see Ref. \Citep{key-56,key-57}),
however, it is fortunate that non determining the normalization constants
throughout this manuscript does not affect the final results. 

We turn now to the particular case where we want to extend the GFVT
for the KGO. In general, we need to perform a substitution of the
momentum operator in Eq. (\eqref{eq:22-1}). Consequently, it is possible
to rewrite Eq. (\eqref{eq:30}) as follows
\begin{equation}
\mathcal{T}=\frac{1}{\sqrt{-g}}\left(\frac{\partial}{\partial r}-m\omega r\right)\left(-\sqrt{-g}\right)\left(\frac{\partial}{\partial r}+m\omega r\right)-\frac{1}{\alpha^{2}r^{2}}\frac{\partial^{2}}{\partial\varphi^{2}}+m^{2},\label{eq:29}
\end{equation}
Similarly, a straightforward calculation based on the procedure that
was carried out in the above discussion, one can obtain the following
differential equation
\begin{equation}
\left[\frac{d^{2}}{dr^{2}}+\frac{1}{r}\frac{d}{dr}+m^{2}\omega^{2}r^{2}-\frac{\sigma^{2}}{r^{2}}+\delta\right]\psi\left(r\right)=0,\label{eq:31-1}
\end{equation}
with
\begin{equation}
\sigma^{2}=\left(\frac{j}{\alpha}\right)^{2},\qquad\delta=E^{2}-m^{2}+2m\omega.\label{eq:41-1}
\end{equation}
Eq. (\eqref{eq:31-1}) is the KGO for spin-0 particle in the space-time
of a static cosmic string. To obtain the solution of this equation,
we first propose a transformation of the radial coordinate 
\begin{equation}
\chi=m\omega r^{2},
\end{equation}
substituting the expression for $\chi$ into Eq. (\eqref{eq:31-1}),
we obtain
\begin{equation}
\left[\frac{d^{2}}{d\chi^{2}}+\frac{1}{\chi}\frac{\partial}{d\chi}-\frac{\sigma^{2}}{4\chi^{2}}+\frac{\delta}{4m\omega\chi}-\frac{1}{4}\right]\psi\left(\chi\right)=0.\label{eq:33-1}
\end{equation}
Now, if we study the asymptotic behavior of the wave function at the
origin and infinity, and since we are looking for regular solutions,
we may suppose a solution that has the form
\begin{equation}
\psi\left(\chi\right)=\chi^{\frac{\left|\sigma\right|}{2}}e^{-\frac{\chi}{2}}F\left(\chi\right),
\end{equation}
As before, this can be substituted back into Eq. (\eqref{eq:33-1}),
then we have
\begin{equation}
\chi\frac{d^{2}F\left(\chi\right)}{d\chi^{2}}+\left(|\sigma|+1-\chi\right)\frac{dF\left(\chi\right)}{d\chi}-\left(\frac{|\sigma|}{2}-\frac{\delta}{4m\omega}+\frac{1}{2}\right)F\left(\chi\right)=0,\label{eq:34}
\end{equation}
This is the confluent hyper-geometric equation \Citep{key-58} whose
solutions are expressed in terms of the confluent hyper-geometric
function type
\begin{equation}
F\left(\chi\right)=_{1}F_{1}\left(\frac{\left|\sigma\right|}{2}-\frac{\delta}{4m\omega}+\frac{1}{2},+1,\chi\right),\label{eq:46-1}
\end{equation}
We should note that the solution (\eqref{eq:46-1}) must be a polynomial
function of degree $n$. However, taking $n\rightarrow\infty$ imposes
a divergence issue. We can have a finite polynomial only if the factor
of the last term in Eq. (\eqref{eq:34}) is a negative integer, meaning,
\begin{equation}
\frac{\left|\sigma\right|}{2}-\frac{\delta}{4m\omega}+\frac{1}{2}=-n\qquad,n=0,1,2,\cdots.
\end{equation}
\begin{figure}[H]
\centering{}\subfloat[$j=1.\,a=0.1,\,\omega=m=1$]{\begin{centering}
\includegraphics[scale=0.4]{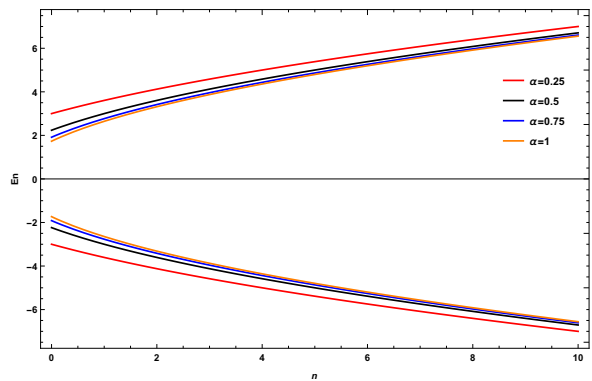}
\par\end{centering}
}\caption{\protect\label{fig.1}Energy of KGO in rotating cosmic string as a
function of the quantum number $n$ for different values of $\alpha$
in (a) and of $a$ in (b).}
\end{figure}

Exploiting this result, and by the insertion of the parameters (\eqref{eq:41-1}),
we can obtain the quantified energy spectrum of KGO in the static
cosmic string space-time, hence,
\begin{equation}
E\left(n,\alpha,j\right)=\pm\sqrt{4m\omega n+\frac{2m\omega\left|j\right|}{\alpha}+m^{2}},\label{eq:48-1}
\end{equation}
We can observe that the energy depends explicitly on the angular deficit
$\alpha$. In other words, the curvature of the space-time which is
influenced by the topological defect i.e, the cosmic string through
the deficit parameter will affects on the relativistic dynamics of
the scalar particle by generating a gravitational field due to the
presence of the wedge angle. Fig. \eqref{fig.1} shows the energy
levels of the KGO as a function of the quantum number $n$ in a static
cosmic string space characterized by a wedge parameter $\alpha$ which
takes different values.

The corresponding wave function is given by
\begin{equation}
\psi\left(t,r,\varphi\right)=|\tilde{C}|\left(m\omega r^{2}\right)^{\frac{\left|j\right|}{2\alpha}}e^{-\frac{m\omega r^{2}}{2}}{}_{1}F_{1}\left(\frac{\left|j\right|}{2\alpha}-\frac{\delta}{4m\omega}+\frac{1}{2},\frac{|j|}{\alpha}+1,m\omega r^{2}\right),\label{eq:35}
\end{equation}
Then the general eigenfunctions are expressed as
\begin{equation}
\psi\left(t,r,\varphi\right)=|\tilde{C}|\left(\begin{array}{c}
1+\frac{E}{\mathcal{N}}\\
1-\frac{E}{\mathcal{N}}
\end{array}\right)\left(m\omega r^{2}\right)^{\frac{\left|j\right|}{2\alpha}}e^{-\frac{m\omega r^{2}}{2}}e^{-i\left(Et-j\varphi\right)}{}_{1}F_{1}\left(\frac{\left|j\right|}{2\alpha}-\frac{\delta}{4m\omega}+\frac{1}{2},\frac{|j|}{\alpha}+1,m\omega r^{2}\right),
\end{equation}
where $|\tilde{C}|$ is the normalization constant.
\begin{center}
\begin{figure}[H]
\begin{centering}
\subfloat[$n=0,\,\alpha=0.5$]{\centering{}\includegraphics[scale=0.6]{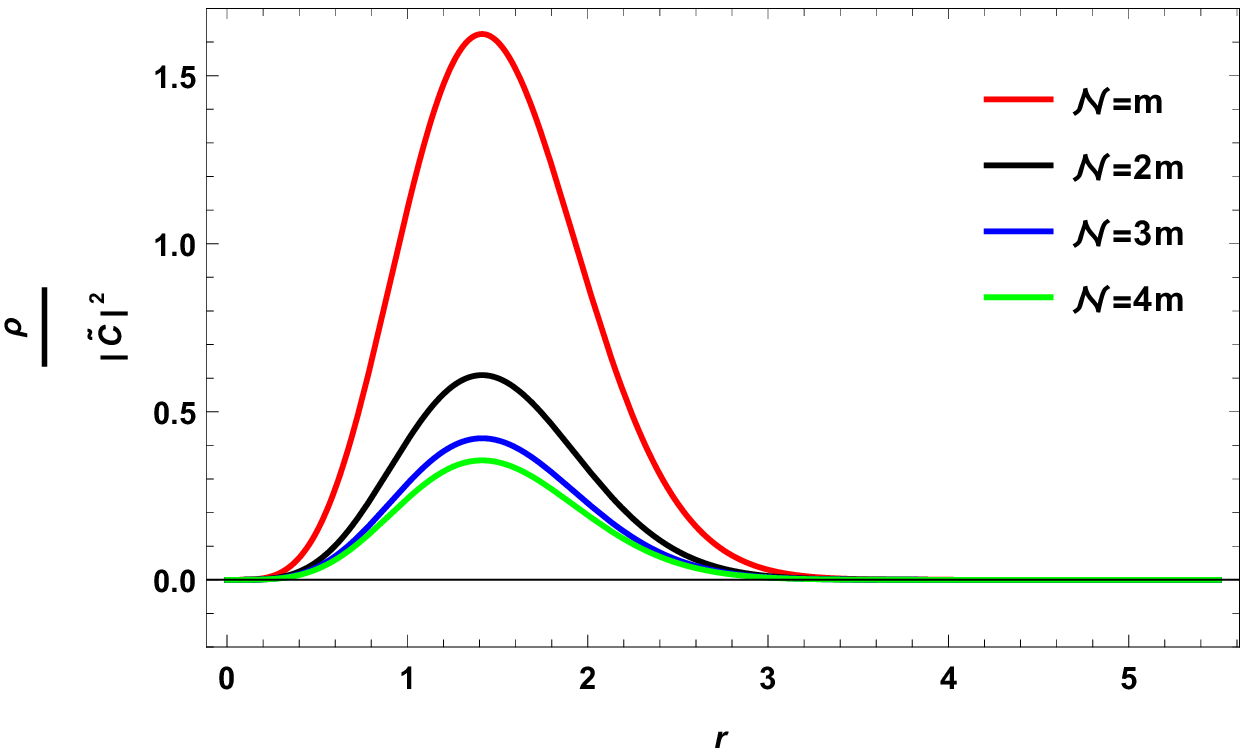}}\subfloat[$n=1,\,\alpha=0.5$]{\centering{}\includegraphics[scale=0.6]{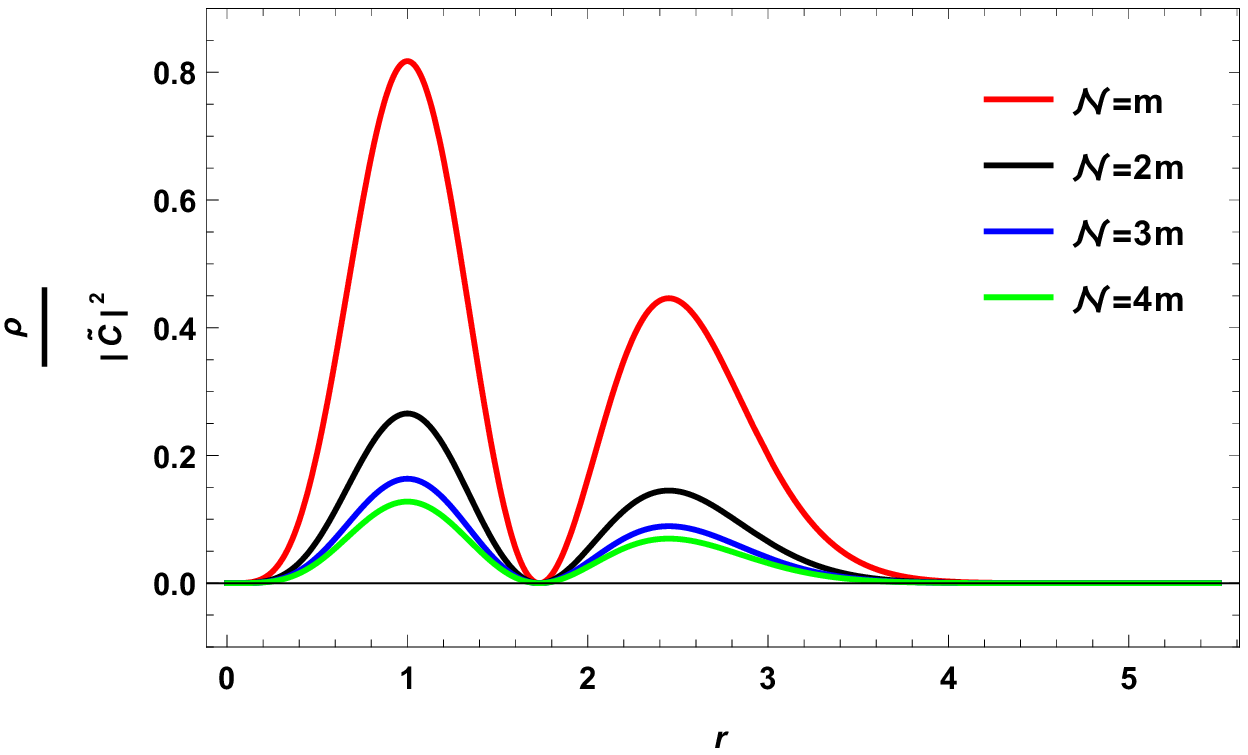}}
\par\end{centering}
\begin{centering}
\subfloat[$n=2,\,\alpha=0.5$]{\centering{}\includegraphics[scale=0.6]{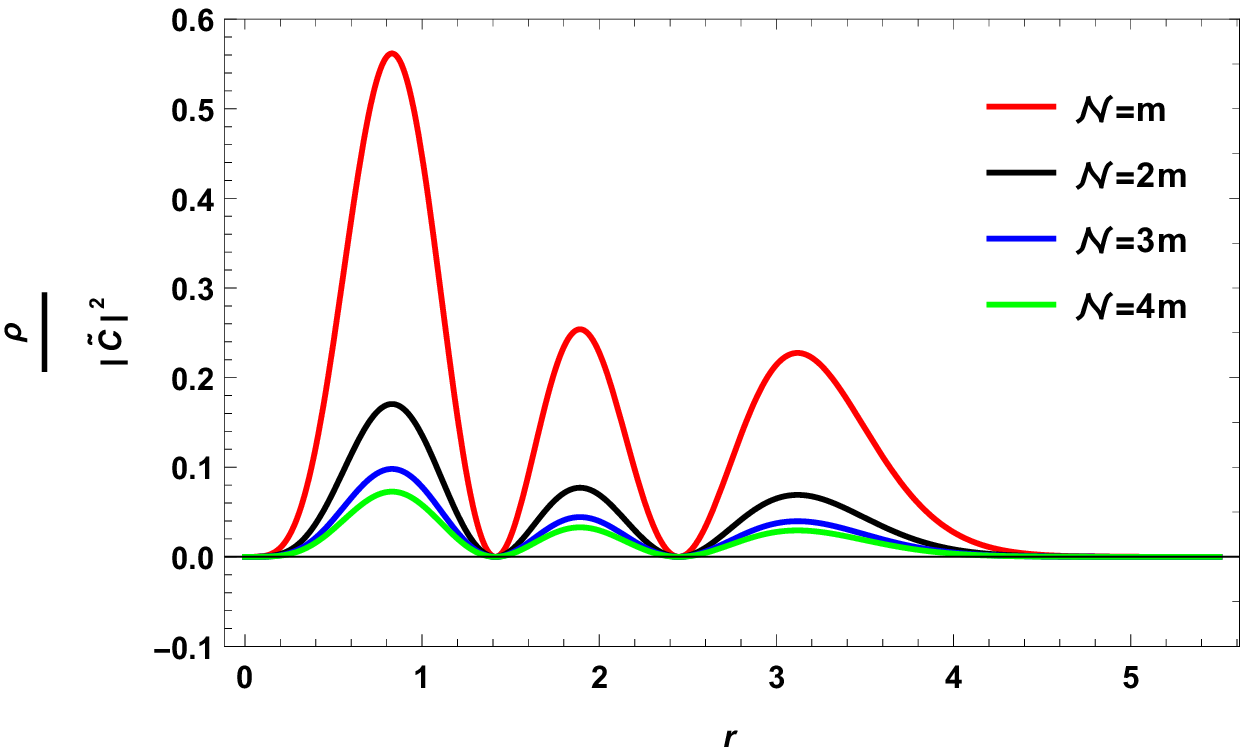}}\subfloat[$n=3,\,\alpha=0.5$]{\centering{}\includegraphics[scale=0.6]{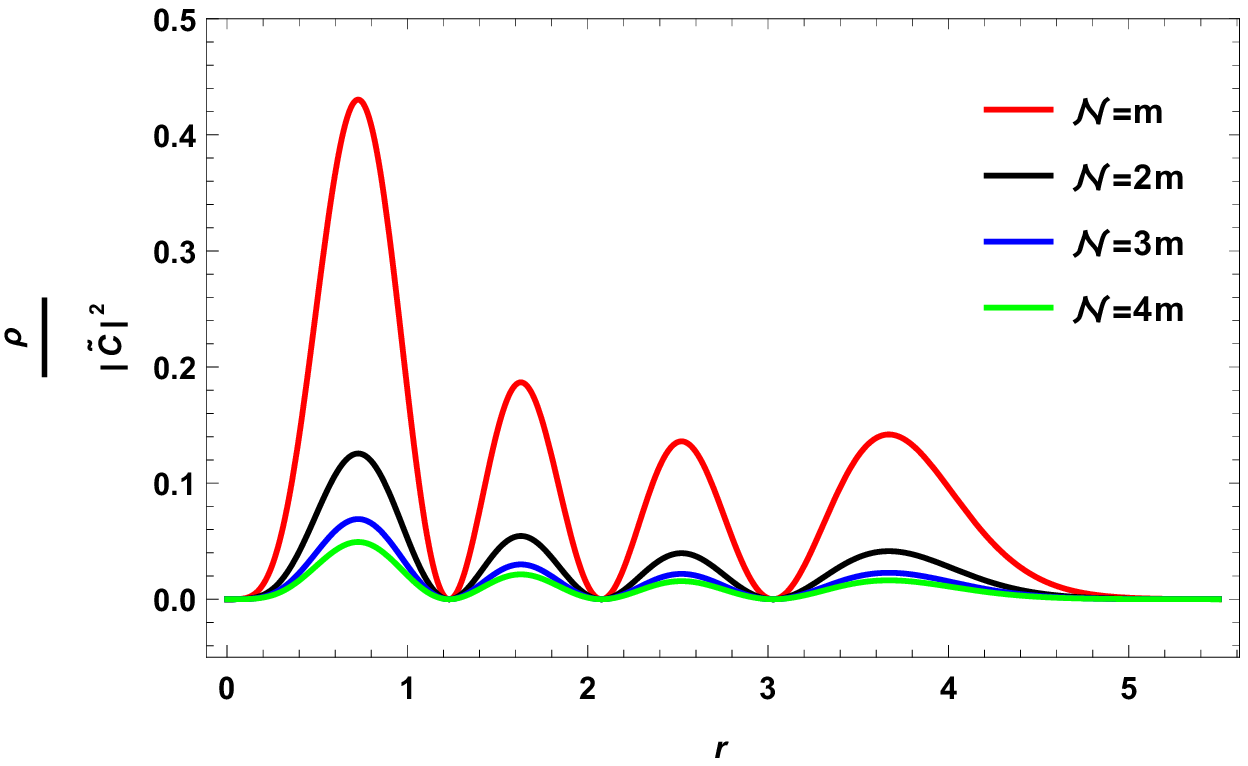}}
\par\end{centering}
\caption{\protect\label{fig:2}Plots of the density of KGO in static cosmic
string as a function of the radial distance $r$ for different values
of $n$ with the parameters $\alpha=0.5$ and $j=1.$}
\end{figure}
\par\end{center}

Having the complete wave-function above, we are in position to calculate
the density $\rho_{_{KGO}}$ corresponds to the KGO given by \citep{key-33}
\begin{equation}
\rho_{_{KGO}}=\psi^{*}\tau_{z}\psi\label{eq:41}
\end{equation}
In Fig. \eqref{fig:2} we plot the density of KGO in static cosmic
string as a function of the radial distance $r.$ Obviously, we can
see that the density is affected by the choice of the variable $\mathcal{N}$. 

\subsection{FV formulation of KGO in rotating cosmic string space-time\eqref{eq:41-1}\protect\label{subsec:3-b}}

In this section we shall analyze the KGO in the background geometry
of a rotating cosmic string in three dimensions. Similarly to the
case studied in Sec. \ref{sec:3}, the equations of motion of a scalar
particle can be achieved by considering the GFVT. Several authors
studied the quantum dynamics of relativistic particles in the space-time
of a cosmic string with rotational effects, and many models were considered
in this context. For instance, in a previous paper by Mazur \Citep{key-59},
the quantum mechanical properties of massive (or massless) particles
in the gravitational field of spinning cosmic strings were discussed.
He showed that the energy should be quantized in the presence of non-zero
angular momentum of the string. Later, Gerbert and Jackiw \Citep{key-60}
have presented solutions for the KG and Dirac equations in the (2+1)-dimensional
space-time created by a massive point particle, with arbitrary angular
momentum. In Ref. \Citep{key-61}, the vacuum expectation value of
the stress-energy tensor for a massless scalar field conformally coupled
to gravitation was discussed. The authors of Ref. \Citep{key-62}
have examined the behavior of a quantum test particle satisfying the
Klein-Gordon equation in a space-time of a spinning cosmic string.
Additionally, it was shown in Ref. \Citep{key-63} that rotating cosmic
string solution of $U(1)$ scalar field theory with a cylindrical
symmetric energy density can be characterized as an extrema of the
field's energy for given angular and linear momenta. Moreover, topological
and geometrical phases due to gravitational field of a cosmic string
that has mass and angular momentum were investigated in Ref. \Citep{key-64}. 

Recently, the subject of gravitational effects of rotating cosmic
strings has attracted considerable interest in connection with the
dynamics of relativistic quantum particles and their properties. For
example, vacuum fluctuations for a massless scalar field around a
spinning cosmic string were studied in Ref. \Citep{key-65} applying
a re-normalization method. In like manner, vacuum polarization of
a scalar field in the gravitational background of a spinning cosmic
string was investigated in Ref. \citep{key-66}. Furthermore, the
authors of Ref. \Citep{key-67} have analyzed the Landau levels of
a spinless massive particle in the spacetime of a rotating cosmic
string by means of a fully relativistic approach. Wang et al. \Citep{key-68}
have treated the model of the KGO coupled to a uniform magnetic field
in the background of the rotating cosmic string. Also, the problem
of a spinless relativistic particle subjected to a uniform magnetic
field in the spinning cosmic string space-time was addressed in Ref.\Citep{key-69}.
Besides,  the relativistic quantum dynamics of a KG scalar field subjected
to a Cornell potential in spinning cosmic string space-time was presented
in Ref. \Citep{key-70}. In addition, the relativistic scalar charged
particle in a rotating cosmic string space-time with Cornell-type
potential and Aharonov-Bohm effect was analyzed in Ref. \Citep{key-29}.
Other publications in which rotating cosmic strings have been considered
to study the quantum properties of relativistic systems could be mentioned
in this respect (e.g, \Citep{key-71,key-72,key-73,key-74,key-75,key-76,key-77,key-78}
and related references therein). Aside from investigating bosonic
particles in that specific space-time, in the literature one can find
various discussions concerning the effects of the topology and geometry
of this space-time on the quantum dynamics of fermionic particles
and their behaviour ( see for instance Refs. \Citep{key-79,key-80,key-81,key-82,key-83,key-84,key-85}). 

In the next discussion, we extend the problem studied in Sec. \ref{sec:3}
to a more general space-time with non-zero angular momentum. We consider
a massive, relativistic spin-0 particle whose wave-function is denoted
by $\Psi$ and satisfies the KG equation \ref{eq:22-1} in the space-time
induced by a (2+1)-dimensional stationary rotating cosmic string which
is described by the following line element \Citep{key-86,key-87,key-59}.
\begin{equation}
ds^{2}=dt^{2}+2a\,dt\,d\phi-dr^{2}-\left(\alpha^{2}r^{2}-a^{2}\right)d\varphi^{2},\label{eq:42}
\end{equation}
always the cylindrical coordinates $(t,r,\varphi)$ are with usual
ranges, and $\alpha=1-4\mu$ is the wedge parameter with $\alpha<1.$
Here, the rotation parameter $a=4J$ has units of distance with $J$
stands for the angular momentum of the string.

The covariant and contravariant components of the metric tensor are
\begin{equation}
g_{\mu\nu}=\left(\begin{array}{ccc}
1 & 0 & a\\
0 & -1 & 0\\
a & 0 & -\left(\alpha^{2}r^{2}-a^{2}\right)
\end{array}\right),\quad g^{\mu\nu}=\left(\begin{array}{ccc}
1-\frac{a^{2}}{\alpha^{2}r^{2}} & 0 & \frac{a}{\alpha^{2}r^{2}}\\
0 & -1 & 0\\
\frac{a}{\alpha^{2}r^{2}} & 0 & -\frac{1}{\alpha^{2}r^{2}}
\end{array}\right).\label{eq:43}
\end{equation}
To study the relativistic quantum motion of a scalar boson interacting
with the gravitational field of the background geometry defined by
the metric \ref{eq:42}, we need to write the generic KG equation
in the curved space in question, given by
\begin{equation}
\left(\square+m^{2}-\xi R\right)\Psi(t,r,\varphi)=0,\label{eq:55-1}
\end{equation}
where $\Psi$ is a two-components vector
\begin{equation}
\Psi(x)=\left(\begin{array}{c}
\phi_{1}\left(x\right)\\
\phi_{2}\left(x\right)
\end{array}\right),\qquad x\equiv(t,r,\varphi).\label{eq:56-2}
\end{equation}
satisfying the condition
\begin{equation}
\psi=\phi_{1}+\phi_{2},\qquad i\mathcal{D}^{\prime}\psi=\mathcal{N}\left(\phi_{1}-\phi_{2}\right),
\end{equation}
with $\mathcal{D}^{\prime}=\frac{\partial}{\partial t}+\mathcal{Y^{\prime}}.$

The next step is to develop the approach used in Sec \ref{sec:3}
for the case of rotating cosmic strings. We adopt the GFVT to derive
the equations of motion for this problem, then we solve them to obtain
the wave functions and the energy spectra. Following Ref. \Citep{key-52,key-53}
, the identification of the Hamiltonian leads to the idea of employing
a two-components formulation of the KG type fields. In order to rewrite
Eq. \eqref{eq:55-1} in the Hamiltonian form, one requires to introduce
new definitions for the quantities presented in Eqs. (\eqref{eq:20}),
(\eqref{eq:30}) and (\eqref{eq:25-1}). In the case under consideration,
we have
\begin{equation}
\mathcal{H}_{GFVT}^{\prime}=\tau_{z}\left(\frac{\mathcal{N}^{2}+\mathcal{T^{\prime}}}{2\mathcal{N}}\right)+i\tau_{y}\left(\frac{-\mathcal{N}^{2}+\mathcal{T^{\prime}}}{2\mathcal{N}}\right)-i\mathcal{Y}^{\prime},\label{eq:56-1}
\end{equation}
\begin{align}
\mathcal{T}^{\prime} & =\partial_{i}\frac{G^{ij}}{g^{00}}\partial_{j}+\frac{m^{2}-\xi R}{g^{00}}+\frac{1}{\mathcal{F}}\nabla_{i}\left(\sqrt{-g}G^{ij}\right)\nabla_{j}\left(\frac{1}{\mathcal{F}}\right)+\sqrt{\frac{\sqrt{-g}}{g^{00}}}G^{ij}\nabla_{i}\nabla_{j}\left(\frac{1}{\mathcal{F}}\right)+\frac{1}{4\mathcal{F}^{4}}\left[\nabla_{i}\left(\mathcal{U}^{i}\right)\right]^{2}\nonumber \\
 & \qquad-\frac{1}{2\mathcal{F}^{2}}\nabla_{i}\left(\frac{g^{0i}}{g^{00}}\right)\nabla_{j}\left(\mathcal{U}^{i}\right)-\frac{g^{0i}}{2g^{00}\mathcal{F}^{2}}\nabla_{i}\nabla_{j}\left(\mathcal{U}^{i}\right),\label{eq:57-1}
\end{align}
\begin{equation}
\mathcal{Y}^{\prime}=\frac{1}{2}\left\{ \partial_{i},\sqrt{-g}\frac{g^{0i}}{g^{00}}\right\} ,\label{eq:58-1}
\end{equation}
where
\begin{equation}
G^{ij}=g^{ij}-\frac{g^{0i}g^{0j}}{g^{00}},\qquad\mathcal{F}=\sqrt{g^{00}\sqrt{-g}},\qquad\mathcal{U}^{i}=\sqrt{-g}g^{0i}.\label{eq:59-1}
\end{equation}
It has been shown in Ref. \Citep{key-53} that the transformations
(\eqref{eq:56-1}), (\eqref{eq:57-1}) and \eqref{eq:58-1} with the
definitions (\eqref{eq:59-1}) are exact and covers any inertial and
gravitational fields. It should be noted that these exact transformations
ensures to obtain the block-diagonal form of the Hamiltonian $\mathcal{H}_{GFVT}$
which does not depend on the parameter $\mathcal{N}.$ For the geometry
(\eqref{eq:43}), the eigenvalues of the operator $\mathcal{Y}^{\prime}$
are given by
\begin{equation}
\mathcal{Y}^{\prime}\Psi\left(t,r,\varphi\right)=\frac{1}{2}\left\{ \partial_{2},\sqrt{-g}\frac{g^{02}}{g^{00}}\right\} \Psi\left(t,r,\varphi\right)=\frac{g^{02}}{g^{00}}\frac{\partial}{\partial\varphi}\Psi\left(t,r,\varphi\right),\label{eq:45}
\end{equation}
where the field $\Psi$ obeys the non-unitary transformation $\Psi\equiv\Phi^{\prime}=\mathcal{F}\Phi$
which permits to obtain pseudo-Hermitian Hamiltonian $\mathcal{H}_{GFVT}^{\prime}=\mathcal{F}\mathcal{H}_{GFVT}^{\prime}\mathcal{F}^{-1},$
$\mathcal{H}_{GFVT}^{\prime}=\tau_{z}\left(\mathcal{H}_{GFVT}^{\prime}\right)^{\dagger}\tau_{z}.$

After some mathematical calculations, Eq. (\eqref{eq:57-1}) takes
the form
\begin{equation}
\mathcal{T}^{\prime}=-\frac{1}{\mathcal{F}}\left[\left(\frac{\partial}{\partial r}\right)\left(\sqrt{-g}\frac{\partial}{\partial r}\right)+\frac{\sqrt{-g}}{\alpha^{2}r^{2}}\frac{\partial^{2}}{\partial\varphi^{2}}\right]\frac{1}{\mathcal{F}}+\frac{m^{2}}{g^{00}}-\left(\frac{g^{02}}{g^{00}}\frac{\partial}{\partial\varphi}\right)^{2},\label{eq:46}
\end{equation}
where for the metric (\eqref{eq:42}) the Ricci scalar vanishes i.e,
$R=0$.

Using the two components fields \eqref{eq:56-2} and the Hamiltonian
(\eqref{eq:56-1}), we can express the field equation (\eqref{eq:55-1})
as the Schrodinger's equation
\begin{equation}
\mathcal{H}_{GFVT}^{\prime}\Psi^{\prime}(x)=i\,\frac{d}{dt}\Psi^{\prime}(x),\qquad x\equiv(t,r,\varphi).\label{eq:47}
\end{equation}
To solve this eigenvalue problem, let us consider for the wave function
the Ansatz below
\begin{equation}
\Psi(x)=\mathcal{F}\Phi\left(x\right)=\mathcal{F}\left(\begin{array}{c}
\phi_{1}\left(r\right)\\
\phi_{2}\left(r\right)
\end{array}\right)e^{-\left(iEt-j\varphi\right)},\label{eq:48}
\end{equation}
substituting (\eqref{eq:48}) into Eq. (\eqref{eq:47}), we find the
following coupled differential equations
\begin{equation}
\begin{aligned}\mathcal{N}^{2}\mathcal{F}\left(\phi_{1}-\phi_{2}\right)+\mathcal{T}^{\prime}\mathcal{F}\left(\phi_{1}-\phi_{2}\right)+2\mathcal{N}\left(\frac{g^{02}}{g^{00}}j\right)\mathcal{F}\phi_{1} & =2\mathcal{N}E\mathcal{F}\phi_{1}\\
\mathcal{N}^{2}\mathcal{F}\left(\phi_{1}-\phi_{2}\right)-\mathcal{T}^{\prime}\mathcal{F}\left(\phi_{1}+\phi_{2}\right)+2\mathcal{N}\left(\frac{g^{02}}{g^{00}}j\right)\mathcal{F}\phi_{2} & =2\mathcal{N}E\mathcal{F}\phi_{2},
\end{aligned}
\label{eq:65-1}
\end{equation}
We can add and subtract the two equations of (\eqref{eq:65-1}) to
get, respectively,
\begin{equation}
\begin{aligned}\mathcal{N}\mathcal{F}\left(\phi_{1}-\phi_{2}\right)+\left(\frac{g^{02}}{g^{00}}j\right)\mathcal{F}\left(\phi_{1}+\phi_{2}\right) & =E\mathcal{F}\left(\phi_{1}+\phi_{2}\right)\end{aligned}
,
\end{equation}
and
\begin{equation}
-\mathcal{T}^{\prime}\mathcal{F}\left(\phi_{1}+\phi_{2}\right)+\mathcal{N}\left(\frac{g^{02}}{g^{00}}j\right)\mathcal{F}\left(\phi_{1}-\phi_{2}\right)=\mathcal{N}E\mathcal{F}\left(\phi_{1}-\phi_{2}\right),\label{eq:50}
\end{equation}
where we have used the relation
\begin{equation}
\phi_{1}=\frac{1}{\mathcal{N}}\left(E-\frac{g^{02}}{g^{00}}j\right)\phi_{2},
\end{equation}
After simple algebraic manipulations we arrive at the following second
order differential equation for the radial function $\psi(r)$
\begin{equation}
\left[\frac{1}{\sqrt{-g}}\frac{\partial}{\partial r}\left(\sqrt{-g}\frac{\partial}{\partial r}\right)-\frac{\left(aE+j\right)^{2}}{\alpha^{2}r^{2}}+E^{2}-m^{2}\right]\psi\left(r\right)=0,\label{eq:53}
\end{equation}
setting $\varsigma^{2}=\frac{\left(aE+j\right)^{2}}{\alpha^{2}},\gamma=E^{2}-m^{2},$
yields
\begin{equation}
\left[\frac{d^{2}}{dr^{2}}+\frac{1}{r}\frac{d}{dr}-\frac{\varsigma^{2}}{r^{2}}+\gamma\right]\varphi_{s}\left(r\right)=0,\label{eq:54}
\end{equation}
Eq. (\eqref{eq:54}) is expressed as a Bessel differential equation,
its solutions can be written in terms of Bessel function of first
kind as
\begin{equation}
\psi\left(r\right)=A^{\prime}\,J_{\frac{aE+j}{\alpha}}\left(\sqrt{E^{2}-m^{2}}\,r\right),\label{eq:55}
\end{equation}
where $A^{\prime}$ is an integration constant. 

The complete eigenstates are given by
\begin{equation}
\psi\left(t,r,\varphi\right)=|C^{\prime}|\left(\begin{array}{c}
1+\frac{E}{\mathcal{N}}\\
1-\frac{E}{\mathcal{N}}
\end{array}\right)e^{-i\left(Et-j\varphi\right)}\,J_{\frac{aE+j}{\alpha}}\left(\sqrt{E^{2}-m^{2}}\,r\right),\label{eq:73-1}
\end{equation}
From now on we proceed to study the Klein Gordon oscillator in a rotating
cosmic string space-time . Firstly, we start by considering a scalar
quantum particle embedded in the background gravitational field of
the space-time described by the metric (\eqref{eq:42}). In this way,
we shall introduce a replacement of the momentum operator $p_{i}\longrightarrow p_{i}+im\omega x_{i}$
where $p_{i}=i\nabla_{i}$ in Eq. (\eqref{eq:57-1}). Then, we have
\begin{equation}
\mathcal{T}^{\prime}=-\frac{1}{\mathcal{F}}\left[\left(\frac{\partial}{\partial r}-m\omega r\right)\left(\sqrt{-g}\left(\frac{\partial}{\partial r}+m\omega r\right)\right)+\frac{\sqrt{-g}}{\alpha^{2}r^{2}}\frac{\partial^{2}}{\partial\varphi^{2}}\right]\frac{1}{\mathcal{F}}+\frac{m^{2}}{g^{00}}-\left(\frac{g^{02}}{g^{00}}\frac{\partial}{\partial\varphi}\right)^{2},\label{eq:58}
\end{equation}
Based on the previous analyses, we shall apply the GFVT for the case
of KGO in the concerned space through the same steps used before.
Inserting Eq. (\eqref{eq:58}) and Eq. (\eqref{eq:45}) into the Hamiltonian
(\eqref{eq:56-1}), then assuming the solution (\eqref{eq:48}) gives
two coupled differential equations similar to those of Eq. (\eqref{eq:65-1})
but with different value of $\mathcal{T}^{\prime}$.

Manipulating exactly the same steps before, we obtain the following
radial equation
\begin{equation}
\left[\frac{\partial^{2}}{\partial r^{2}}+\frac{1}{r}\frac{\partial}{\partial r}-m^{2}\omega^{2}r^{2}-\frac{\vartheta^{2}}{r^{2}}+\delta\right]\psi\left(r\right)=0,\label{eq:61}
\end{equation}
where we have defined
\begin{equation}
\vartheta^{2}=\left(\frac{aE+j}{\alpha}\right)^{2},\qquad\delta=E^{2}-m^{2}+2m\omega.\label{eq:75-1}
\end{equation}
To solve Eq. (\eqref{eq:61}), we introduce a new dimensionless variable
$\mathcal{R}=m\omega r^{2}$, and by substitution into Eq. (\ref{eq:61})
, the resulting equation reads
\begin{equation}
\left[\frac{d^{2}}{d\mathcal{R}^{2}}+\frac{1}{\mathcal{R}}\frac{d}{d\mathcal{R}}-\frac{\vartheta^{2}}{4\mathcal{R}^{2}}-\frac{1}{4}+\frac{\delta}{4m\omega\mathcal{R}}\right]\psi\left(\mathcal{R}\right)=0,\label{eq:63}
\end{equation}
To eliminate the term $d\psi(\mathcal{R})\big/d\mathcal{R},$ we consider
the following change of variable
\begin{equation}
\psi(\mathcal{R})=\mathcal{R}^{-\frac{1}{2}}\Xi\left(\mathcal{R}\right),
\end{equation}
Eq. (\eqref{eq:63}) then becomes
\begin{equation}
\frac{d^{2}\Xi\left(\mathcal{R}\right)}{d\mathcal{R}^{2}}+\left[-\frac{1}{4}+\frac{\delta}{4m\omega\mathcal{R}}+\frac{\frac{1}{4}-\left(\frac{\vartheta}{2}\right)^{2}}{\mathcal{R}^{2}}\right]\Xi\left(\mathcal{R}\right)=0,
\end{equation}
and it has the form of the Whittaker differential equation \Citep{key-55}.
The general solution of this equation, which is regular at the origin,
is given by 
\begin{equation}
\Xi\left(\mathcal{R}\right)=|C|\mathcal{R}^{-\tfrac{1}{2}}M\left(\frac{\delta}{4m\omega},\frac{|\vartheta|}{2},\mathcal{R}\right),\label{eq:64}
\end{equation}
where $|C|$ is an arbitrary constant and $M\left(\frac{\delta}{4m\omega},\frac{|\vartheta|}{2},\mathcal{R}\right)$
is the Whittaker M-function defined via the confluent hyper-geometric
functions as
\begin{equation}
M\left(\frac{\delta}{4m\omega},\frac{|\vartheta|}{2},\mathcal{R}\right)=e^{-\frac{\mathcal{R}}{2}}\mathcal{R}^{\frac{\left|\vartheta\right|+1}{2}}{}_{1}F_{1}\left(\frac{|\vartheta|}{2}-\frac{\delta}{4m\omega}+\frac{1}{2},|\vartheta|+1,\mathcal{R}\right),\label{eq:67}
\end{equation}
Using the definition (\eqref{eq:67}), the final expression of the
wave-function of the spinless KGO propagating in the rotating cosmic
string background can be represented as
\begin{equation}
\psi\left(t,r,\varphi\right)=|\mathtt{C}|\left(\begin{array}{c}
1+\frac{E}{\mathcal{N}}\\
1-\frac{E}{\mathcal{N}}
\end{array}\right)\left(m\omega r^{2}\right)^{\frac{\left|\vartheta\right|}{2}}e^{-\frac{m\omega}{2}r^{2}}e^{-i\left(Et-j\varphi\right)}{}_{1}F_{1}\left(\frac{|\vartheta|}{2}-\frac{\delta}{4m\omega}+\frac{1}{2},|\vartheta|+1,m\omega r^{2}\right),\label{eq:68}
\end{equation}
where the parameters $\vartheta$ and $\delta$ are defined in Eq.
(\eqref{eq:75-1}).

Again, The asymptotic behavior of the confluent hyper-geometric function
implies that
\begin{equation}
\frac{|\vartheta|}{2}-\frac{\delta}{4m\omega}+\frac{1}{2}=-n,\label{eq:69}
\end{equation}
\begin{figure}[H]
\centering{}\subfloat[$j=1.\,a=0.1,\,\omega=m=1$]{\begin{centering}
\includegraphics[scale=0.4]{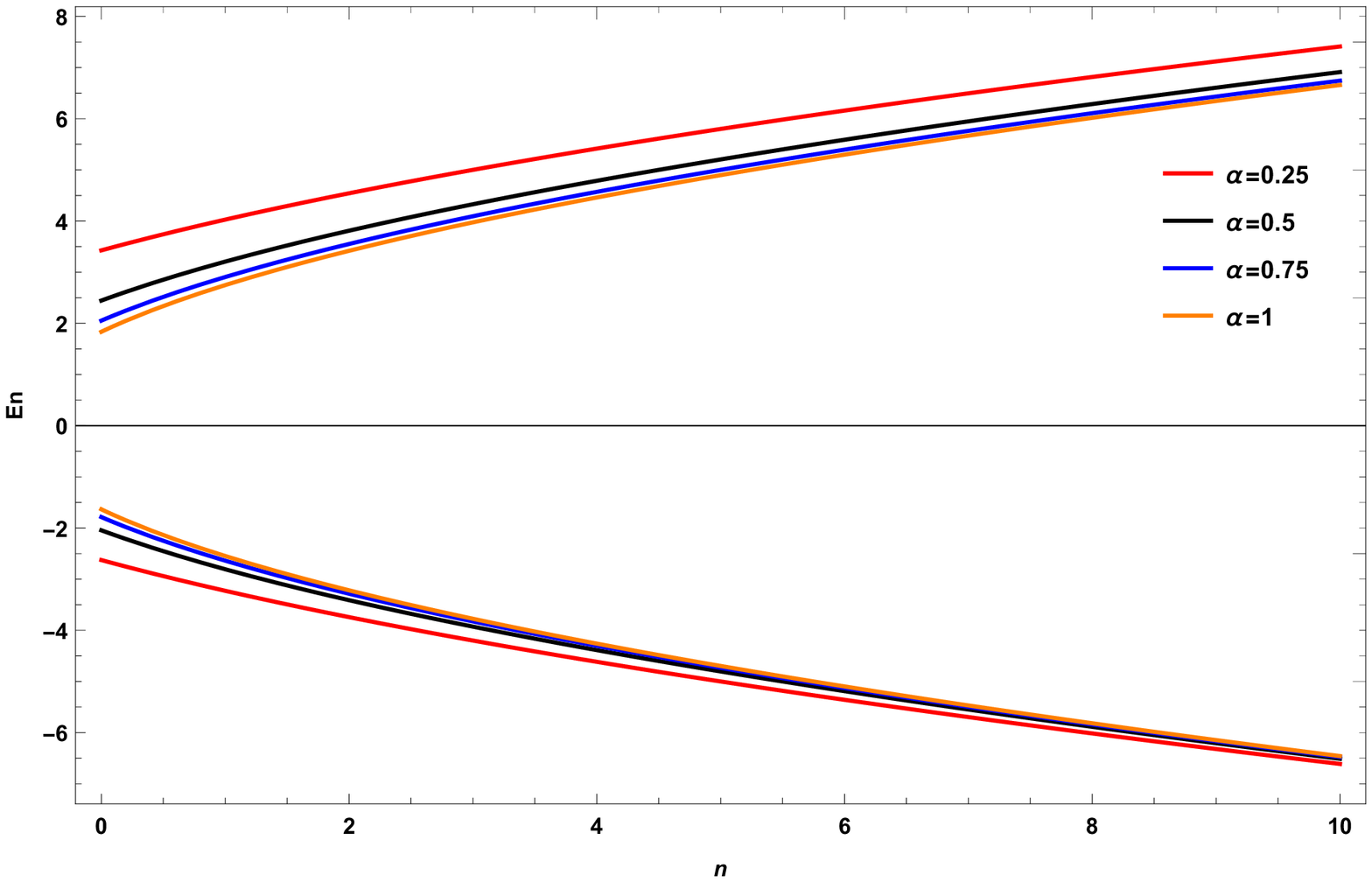}
\par\end{centering}
}\subfloat[$j=1.\,\alpha=0.5,\,\omega=m=1$]{\begin{centering}
\includegraphics[scale=0.4]{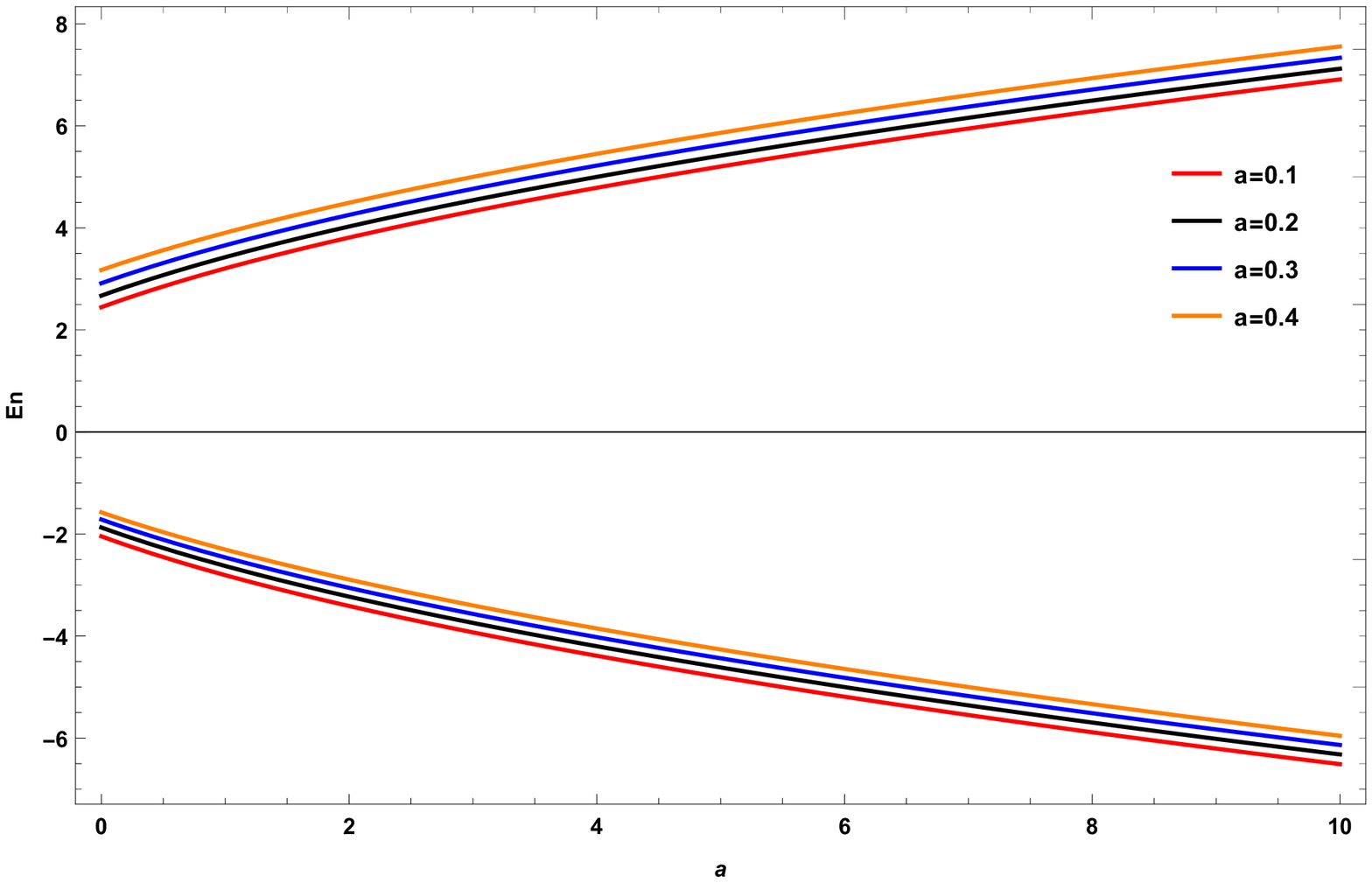}
\par\end{centering}
}\caption{\protect\label{fig:2-1}Energy of KGO in rotating cosmic string as
a function of the quantum number $n$ for different values of $\alpha$
in (a) and of $a$ in (b).}
\end{figure}
 Hence, after inserting $\vartheta$ and $\delta$ and by solving
Eq. (\eqref{eq:69}) for $E$, we obtain the energy levels for our
scalar particle
\begin{equation}
E^{\pm}(n,\alpha,a,j)=\frac{am\omega}{\alpha}\pm\sqrt{4m\omega n+\left(\frac{am\omega}{\alpha}\right)^{2}+\frac{2m\omega|j|}{\alpha}+m^{2},}\label{eq:70}
\end{equation}
\begin{figure}[H]
\begin{centering}
\subfloat[$n=0,\,a=0.1,\,\alpha=0.5$]{\centering{}\includegraphics[scale=0.6]{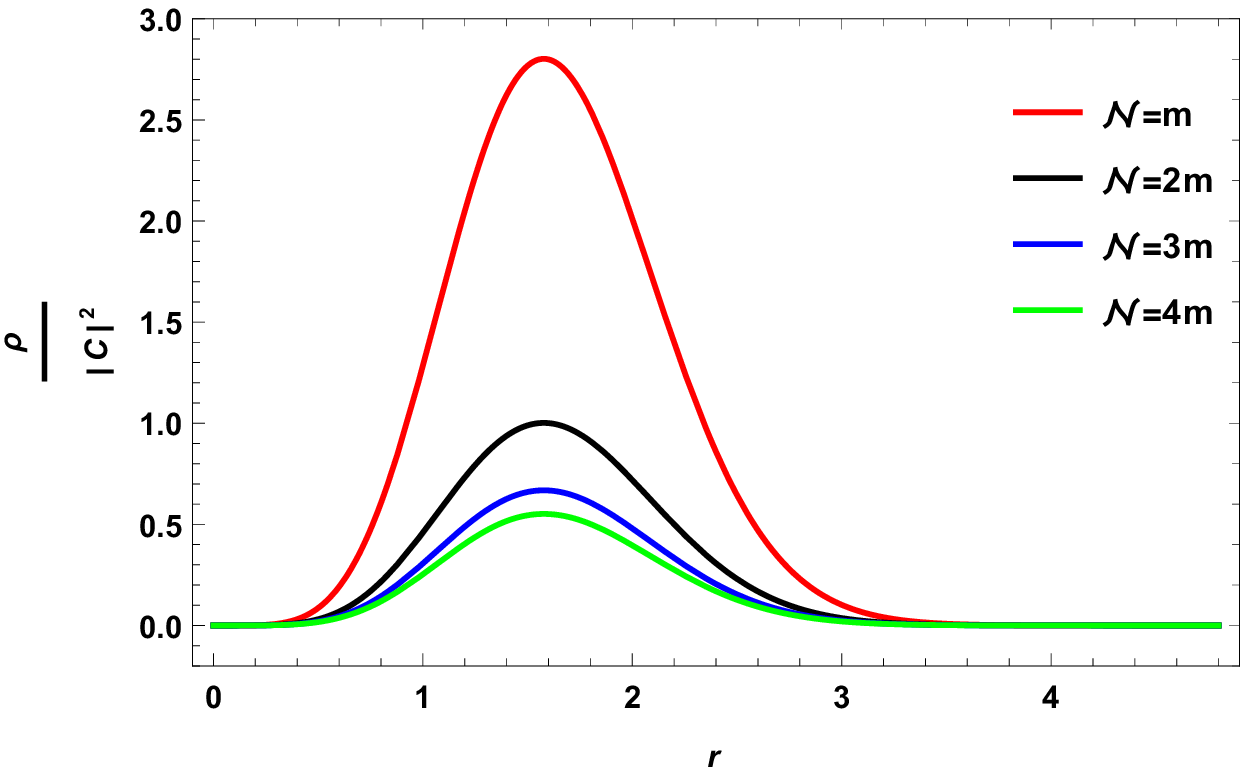}}\subfloat[$n=1,\,a=0.1,\,\alpha=0.5$]{\centering{}\includegraphics[scale=0.6]{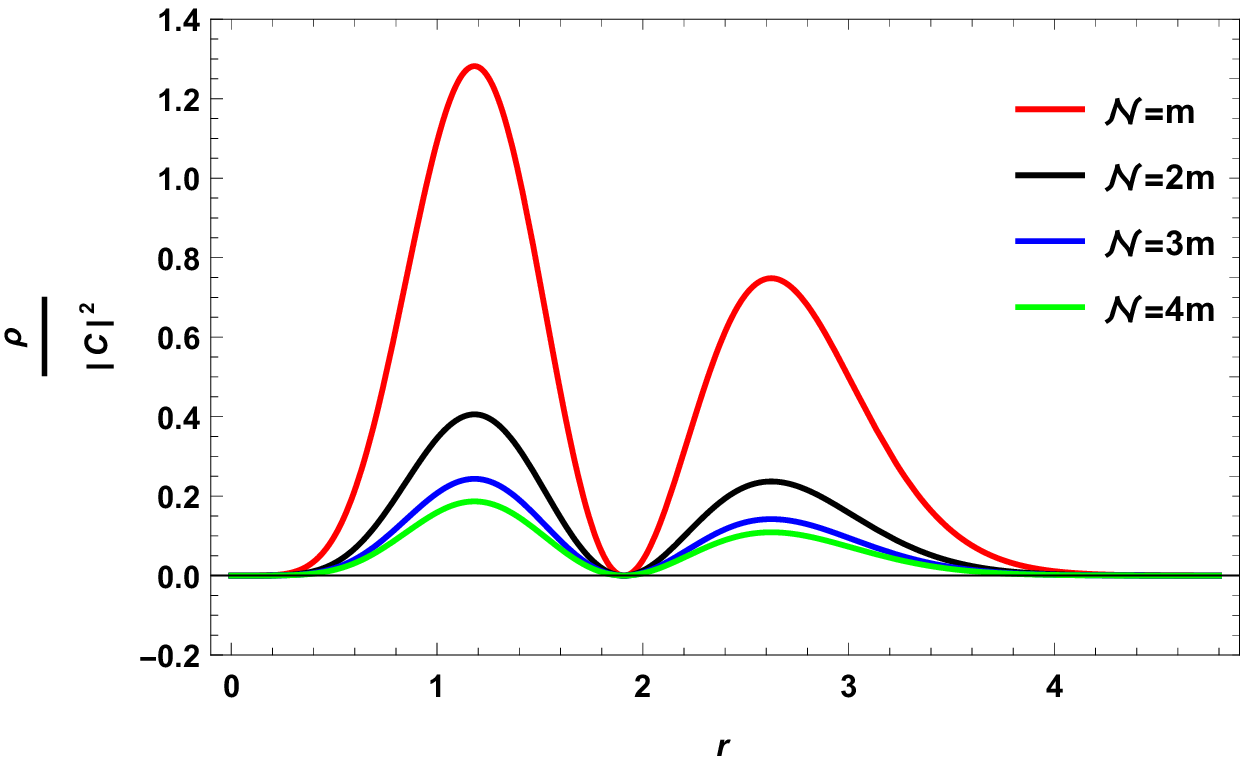}}
\par\end{centering}
\centering{}\subfloat[$n=2,\,a=0.1,\,\alpha=0.5$]{\centering{}\includegraphics[scale=0.6]{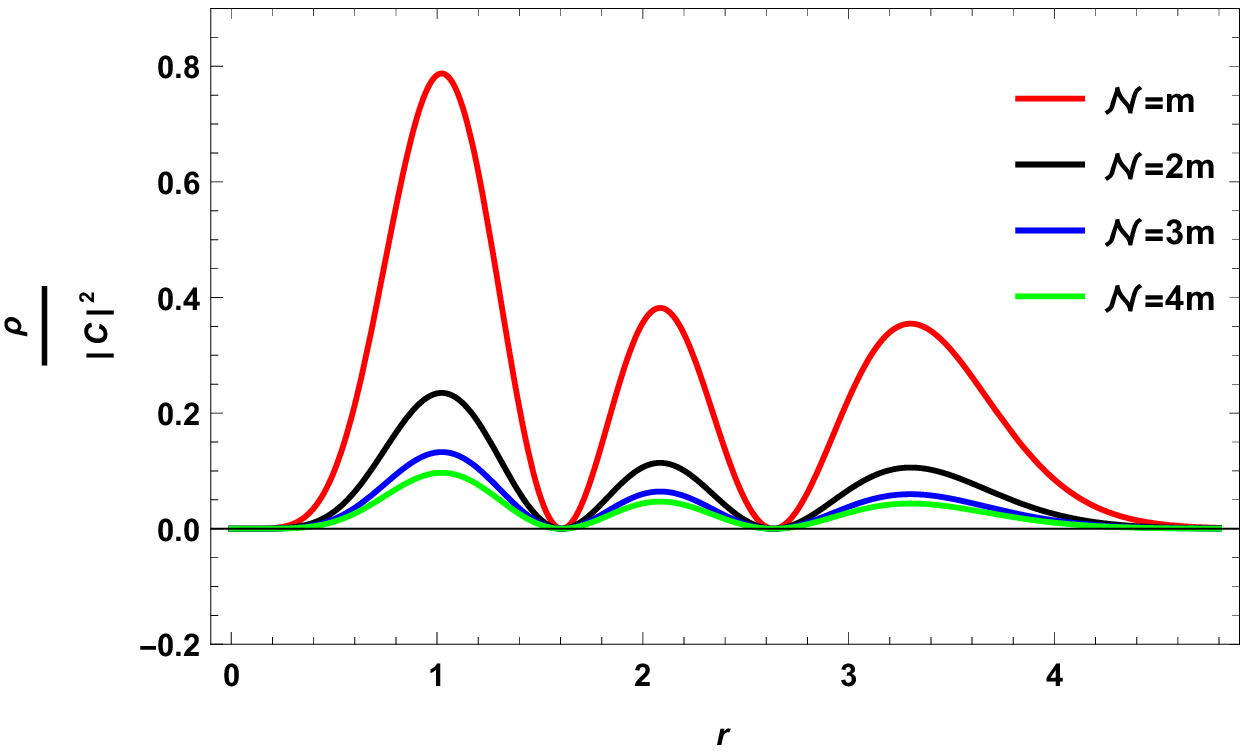}}\subfloat[$n=3,\,a=0.1,\,\alpha=0.5$]{\centering{}\includegraphics[scale=0.6]{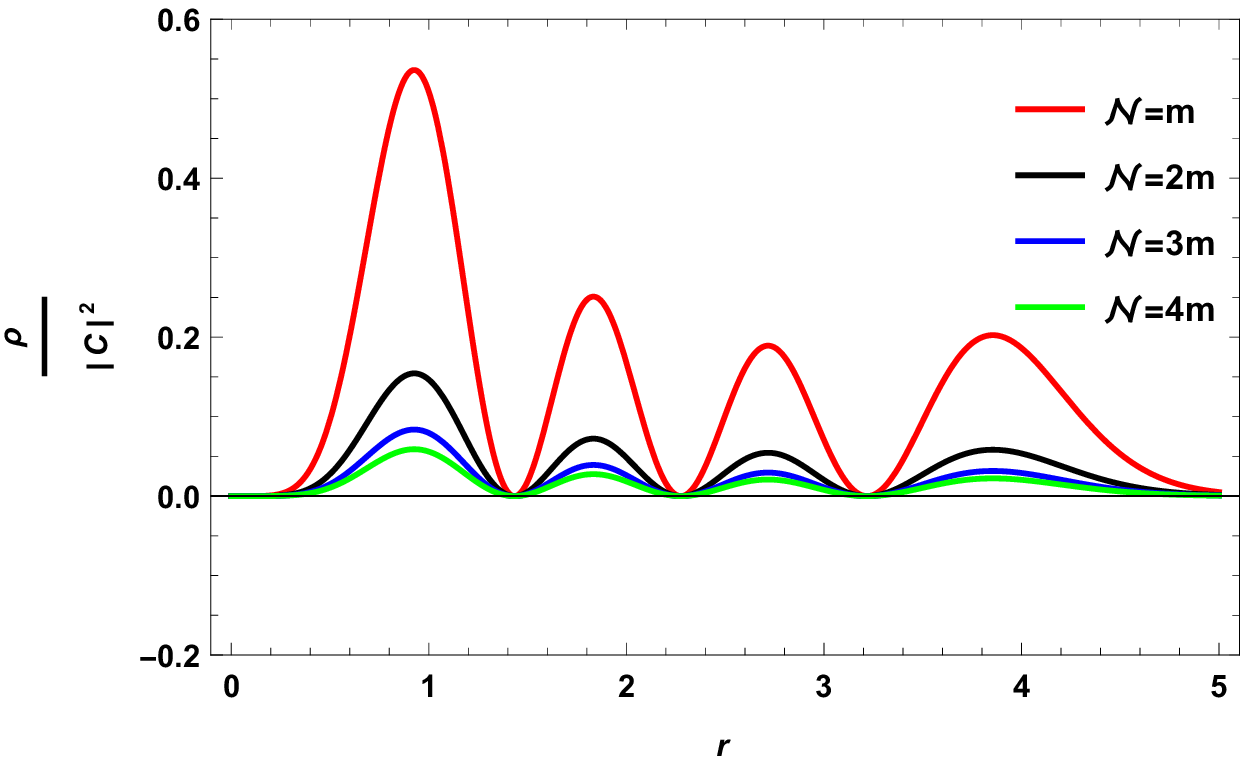}}\caption{\protect\label{fig:3-1}Plots of the density of KGO in rotating cosmic
string as a function of the radial distance $r$ for different values
of $n$ with the parameters $a=0.5$ and $j=1.$}
\end{figure}
\begin{figure}[H]
\begin{centering}
\subfloat[$\mathcal{N}=m,\,a=0.1$]{\centering{}\includegraphics[scale=0.6]{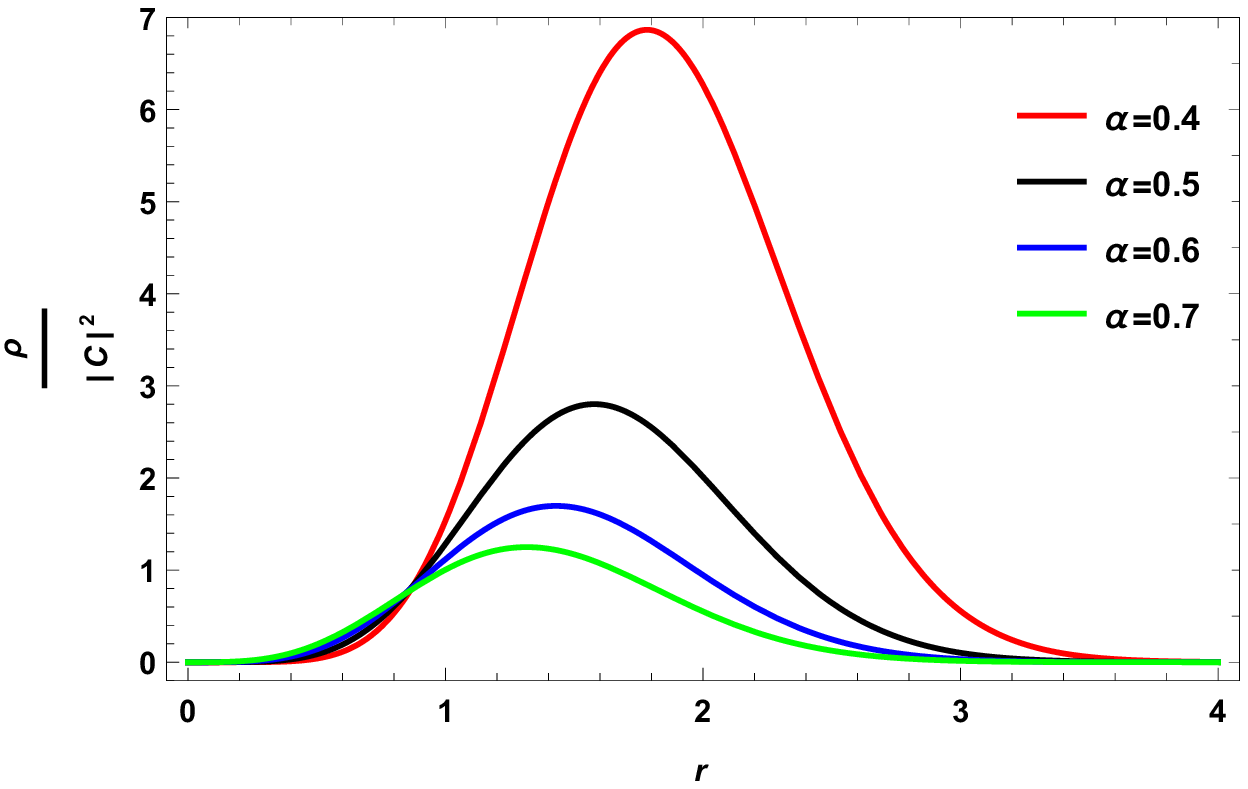}}\subfloat[$\mathcal{N}=m,\,\alpha=0.5$]{\centering{}\includegraphics[scale=0.6]{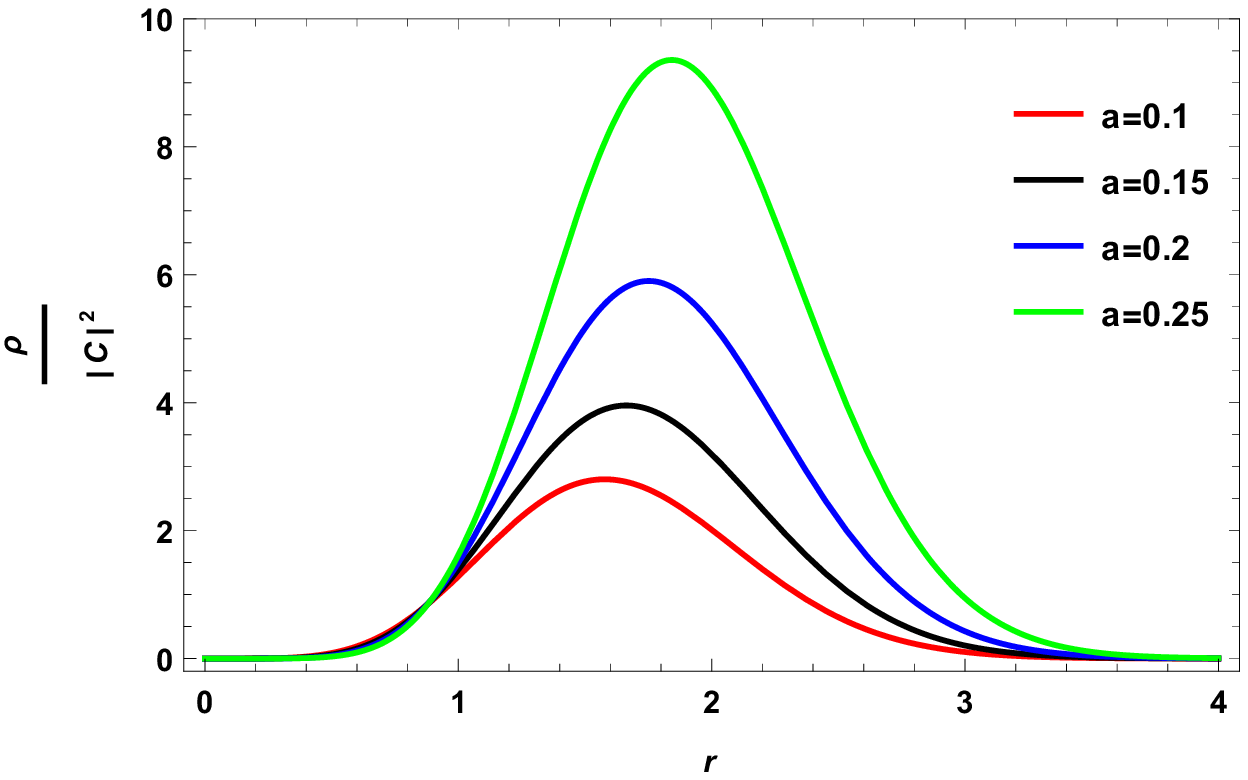}}
\par\end{centering}
\centering{}\caption{\protect\label{fig:3-1-1}Plots of the density of KGO in rotating
cosmic string as a function of the radial distance $r$ for different
values of $\alpha$ and $a$ with the parameters $m=\omega=1$ and
$j=1.$}
\end{figure}

As an interesting side result, we have an expression for the energy
in which we see all the parameters that characterize the background
geometry. Indeed, the energy spectrum does not depend only on the
angular deficit parameter $\alpha$, but also on the rotation parameter
$a$. We expect that the external gravitational field due to a rotating
cosmic string affects on the relativistic dynamics of the quantum
particle, essentially, since it has non-zero curvature concentrated
along the $z$-axis and its angular momentum appears throughout the
parameter $a.$ By comparing the energy (\eqref{eq:70}) with the
energy of KGO in a static cosmic string (\eqref{eq:48-1}), we notice
that $E\left(n,\alpha,a,j\right)$ is shifted by the amount $am\omega\big/\alpha$
and enlarged by a conical factor. We plot the energy eigenvalues as
a function of the quantum number $n$ with fixed parameters $\omega=j=1$
and for different values of the variable $\alpha$ and $a$ shown
in Fig. \eqref{fig:2-1}.

Taking this result into the consideration, we can find the density
for the FV-KGO in the rotating cosmic string space as a function of
the radial distance $r$ and fixed parameter $\alpha$. Fig. \eqref{fig:3-1}shows
that the density is affected when choosing the $\mathcal{N}$ variable
for different situations. Now, the case where $\mathcal{N}=m$ that
corresponds with the case of KGO is presented in the figure. \eqref{fig:3-1-1}.
This figure shows that density is strongly dependent with the parameter
of rotation $a$.

\section{Thermal properties of FVO in Rotating Cosmic String}

This section is devoted to look into the thermal properties which
may arise throughout our study of the KGO in the space-time produced
by a rotating cosmic string. Having already obtained the energy spectrum
of KGO in that particular space, we are now in a position to present
the thermodynamics of the model encountered in Sec. \eqref{subsec:3-b}
by calculating the partition function which provides all the physical
information of our quantum system. To better illustrate our results,
we present several figures in the next discussion.

\subsection{Partition function}

We are interested in determining the partition function of KGO interacting
with the gravitational field induced by rotating cosmic string. Consequently,
the statistical quantities such as the free energy $F,$the mean energy
$U,$the entropy $S$and the specific heat $C_{v}$, associated with
the relativistic spinless particle in that space-time can be obtained
from the partition function. 

The canonical ensemble partition function at finite temperature $T$
is given by
\begin{equation}
Z=\stackrel[n=0]{\infty}{\sum}e^{-\beta E_{n}},\label{eq:73}
\end{equation}
where $E_{n}$ denotes the corresponding energy eigenvalues and $\beta=\frac{1}{K_{B}T}$
with $K_{B}$ being the Boltzmann constant. Given the spectrum \eqref{eq:70}
and by considering only the positive energies i.e, $E_{n}=E_{n}^{+}$,
we can express the partition function as
\begin{equation}
Z=\stackrel[n=0]{\infty}{\sum}e^{-\beta\left(\mathsf{Y}+\sqrt{\mathsf{A}n+\mathsf{B}}\right)}=e^{-\beta\mathsf{Y}}\stackrel[n=0]{\infty}{\sum}e^{-\beta\sqrt{\mathsf{A}n+\mathsf{B}}},
\end{equation}
where
\[
\mathsf{Y}=\frac{am\omega}{\alpha},\qquad\mathsf{A}=4m\omega,\qquad\mathsf{B}=\left(\frac{am\omega}{\alpha}\right)^{2}+\frac{2m\omega|j|}{\alpha}+m^{2}
\]
In what follows, we closely follow the methods outlined in Ref.\Citep{key-88}
and revised in \Citep{key-89} with the intention of approximating
the infinite sum apparent in Eq. \eqref{eq:73}. For this purpose,
we shall begin with rewriting the series representation of the partition
function $Z$ in terms of the Zeta function. Hence we have

\begin{equation}
\stackrel[n=0]{\infty}{\sum}e^{-\beta\sqrt{\mathsf{A}n+\mathsf{B}}}=e^{-\widetilde{\beta}}+\stackrel[n^{\prime}=0]{\infty}{\sum}e^{-\tilde{\beta}\sqrt{\mathsf{D}}\sqrt{\frac{1}{\mathsf{D}}+\left(n^{\prime}+1\right)}},\qquad n=n^{\prime}+1.\label{eq:74}
\end{equation}
where we have introduced the notation
\[
\tilde{\beta}=\beta\sqrt{\mathsf{B}},\qquad\mathsf{D=\frac{A}{B}.}
\]
Using the Cahen-Mellin integral \Citep{key-90,key-91}
\[
e^{-\mathit{z}}=\frac{1}{2\pi i}\stackrel[\mathfrak{c}-i\infty]{\mathfrak{c}+i\infty}{\int}\Gamma\left(s\right)z^{-s}ds,\qquad\left(|\textrm{arg}z|<\frac{\pi}{2};z\neq0\right).
\]
where the path of integration is the vertical line $\Re(s)=\mathfrak{c}$,
with $\mathfrak{c}\in\mathbb{R}^{*}$, lying to the right of all the
poles of $\Gamma(s),$ the term $\stackrel[n^{\prime}=0]{\infty}{\sum}e^{-\tilde{\beta}\sqrt{\mathsf{D}}\sqrt{\frac{1}{\mathsf{D}}+\left(n^{\prime}+1\right)}}$
can be expressed as
\begin{equation}
\stackrel[n^{\prime}=0]{\infty}{\sum}e^{-\tilde{\beta}\sqrt{\mathsf{D}}\sqrt{\frac{1}{\mathsf{D}}+\left(n^{\prime}+1\right)}}=\frac{1}{2\pi i}\stackrel[\mathfrak{c}-i\infty]{\mathfrak{c}+i\infty}{\int}\left(\tilde{\beta}\sqrt{\mathsf{D}}\right)^{-s}\Gamma\left(s\right)\zeta_{H}\left(\frac{s}{2},\frac{1}{\mathsf{D}}+1\right),\label{eq:87}
\end{equation}
\begin{figure}[H]
\centering{}\includegraphics[scale=0.6]{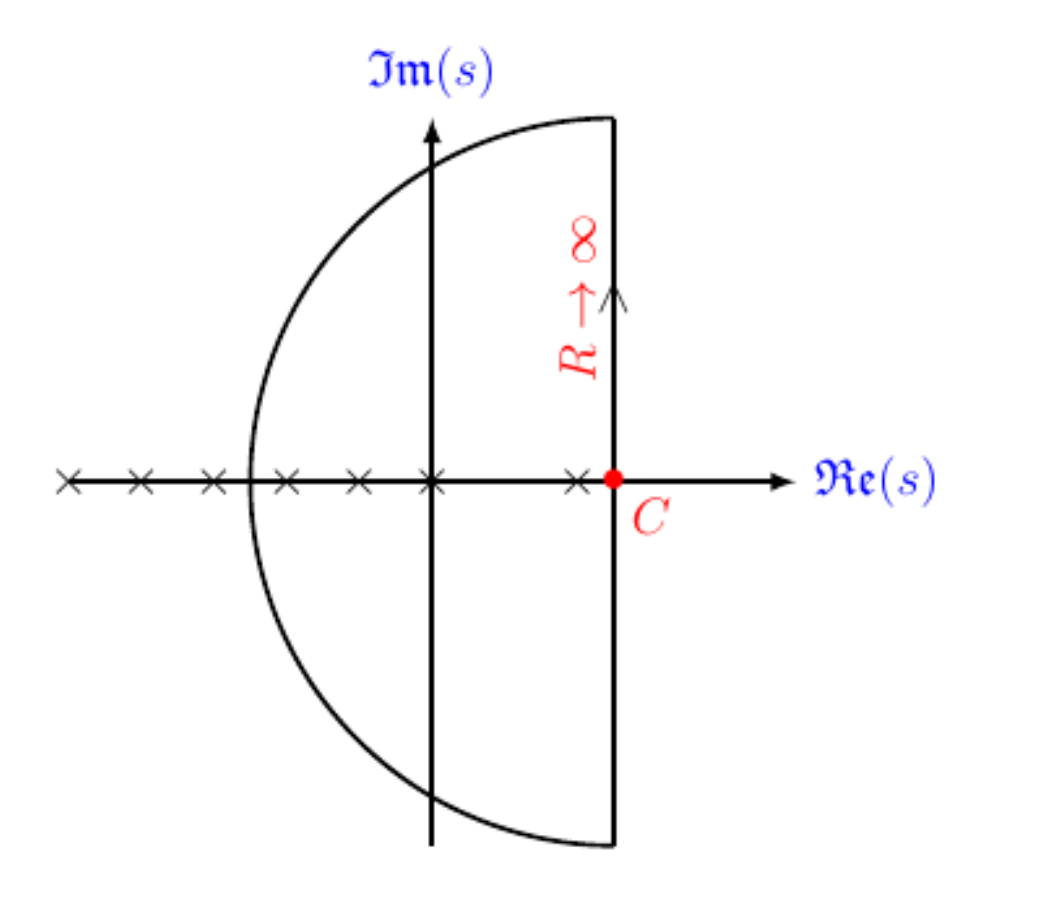}\caption{\protect\label{fig:contour}Contour Integration of the Mellin transformation
of the problems in question \citep{key-89,key-91} }
\end{figure}

$\Gamma(s)$ and $\zeta_{H}\left(\frac{s}{2},\frac{1}{\mathsf{D}}+1\right)$
refers to Euler Gamma and Hurwitz zeta functions, receptively. We
note that the integral converges only when $s\big/2>1$ requires $\mathfrak{c}>2.$
As we need to work out the integral \eqref{eq:87}, it is clear that
the integrand has poles at $s=-k;(k=0,1,2,\cdots)$ of $\Gamma(s),$with
residue $(-1)^{k}\big/k!$ (see Figure. \eqref{fig:contour}). In
addition, we have a simple pole at $s=2$ of $\zeta_{H}\left(\frac{s}{2},\frac{1}{\mathsf{D}}+1\right)$
with residue equals to $2.$

A straightforward evaluation of the residues at the poles $\{s=0,2,-k\}$
leads to the following expression of the partition function
\begin{equation}
Z(\tilde{\beta})=e^{-\tilde{\beta}\mathsf{YB}^{\nicefrac{-1}{2}}}\left[-\frac{1}{2}+\frac{1}{\mathsf{D}}\left(\frac{2}{\widetilde{\beta}^{2}}-1\right)+e^{-\widetilde{\beta}}+\stackrel[n=1]{\infty}{\sum}\frac{\left(-\widetilde{\beta}\sqrt{\mathsf{D}}\right)}{n!}^{n}\zeta_{H}\left(-\frac{n}{2},1+\frac{1}{\mathsf{D}}\right)\right],\label{eq:88-1}
\end{equation}
where we have used the identities 
\begin{equation}
\zeta_{H}\left(0,p\right)=\frac{1}{2}-p.
\end{equation}

\subsection{Thermal properties}

Having estimated the partition function of KGO in rotating cosmic
string, we are allowed to compute the main thermodynamic quantities,
\begin{equation}
F=-\frac{\sqrt{B}}{\tilde{\beta}}\ln\left(Z\right),U=-\sqrt{B}\frac{d\ln\left(Z\right)}{d\tilde{\beta}},\label{eq:88}
\end{equation}
\[
S=\ln\left(Z\right)-\tilde{\beta}\frac{d\ln\left(Z\right)}{d\tilde{\beta}},\,C_{V}=\tilde{\beta}{}^{2}\frac{d^{2}\ln\left(Z\right)}{d\tilde{\beta}^{2}}.
\]
The series representations of the partition function $Z$ in terms
of the $\zeta_{H}$-function and the Boltzmann representation introduced
in Eq. (\eqref{eq:88-1}) can be evaluated numerically considering
only a finite number of terms in the series. 

\begin{figure}[h]
\begin{centering}
\subfloat[Free energy $F$ versus the reduced temperature $\tau$]{\centering{}\includegraphics[scale=0.3]{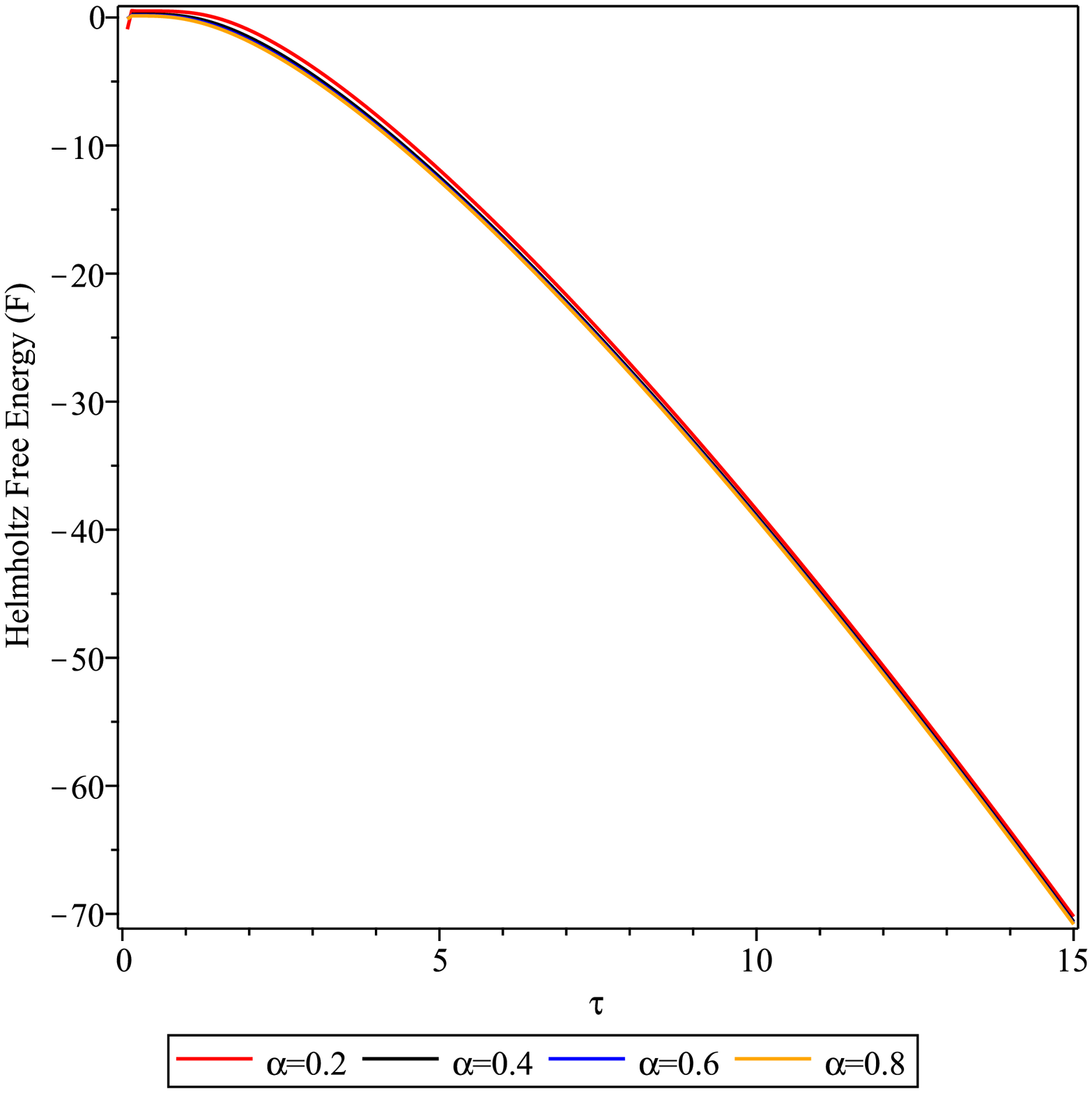}}\subfloat[Total energy $U$ versus the reduced temperature $\tau$]{\centering{}\includegraphics[scale=0.3]{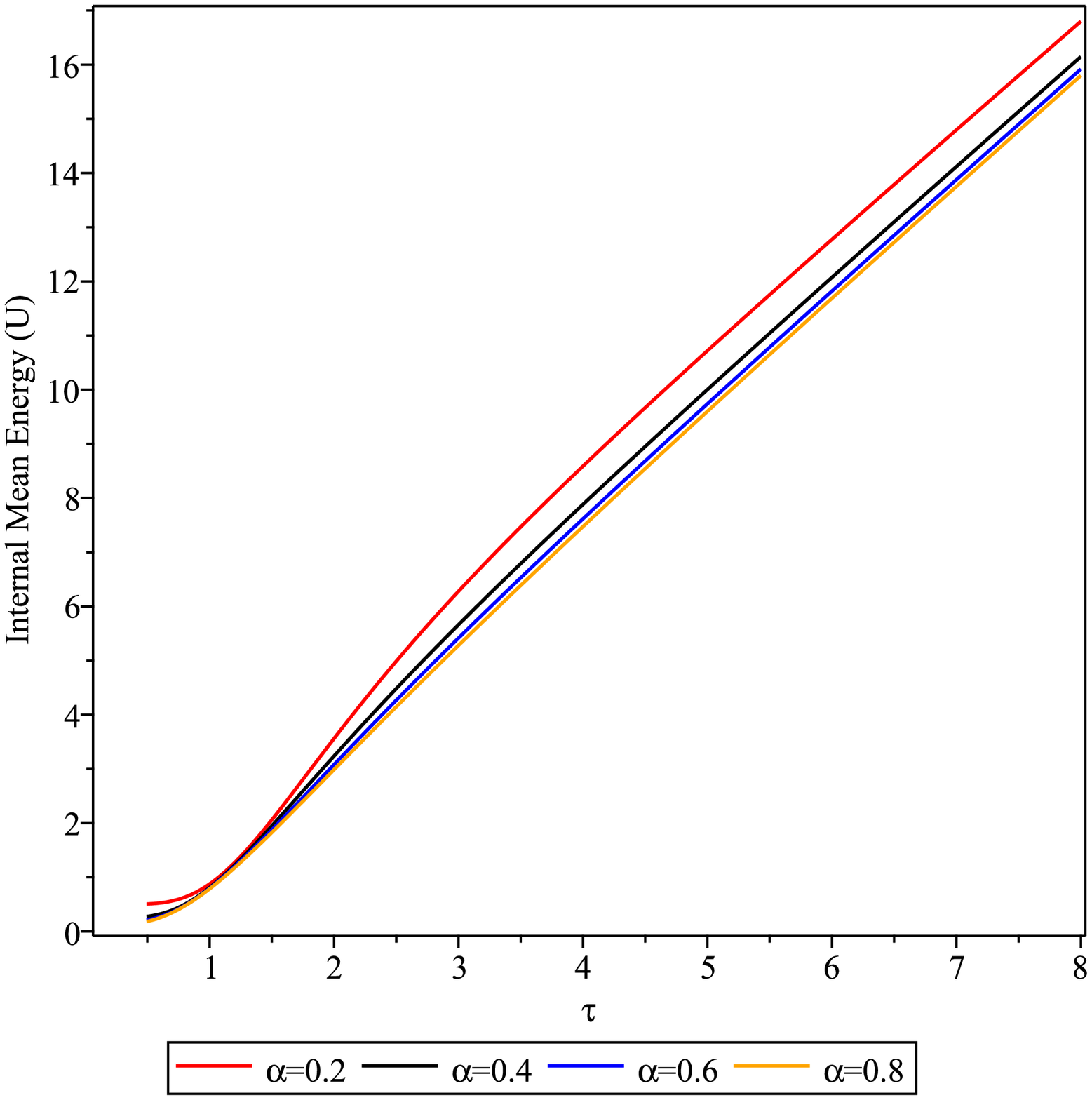}}
\par\end{centering}
\begin{centering}
\subfloat[Entropy $S$ versus the reduced temperature $\tau$]{\centering{}\includegraphics[scale=0.3]{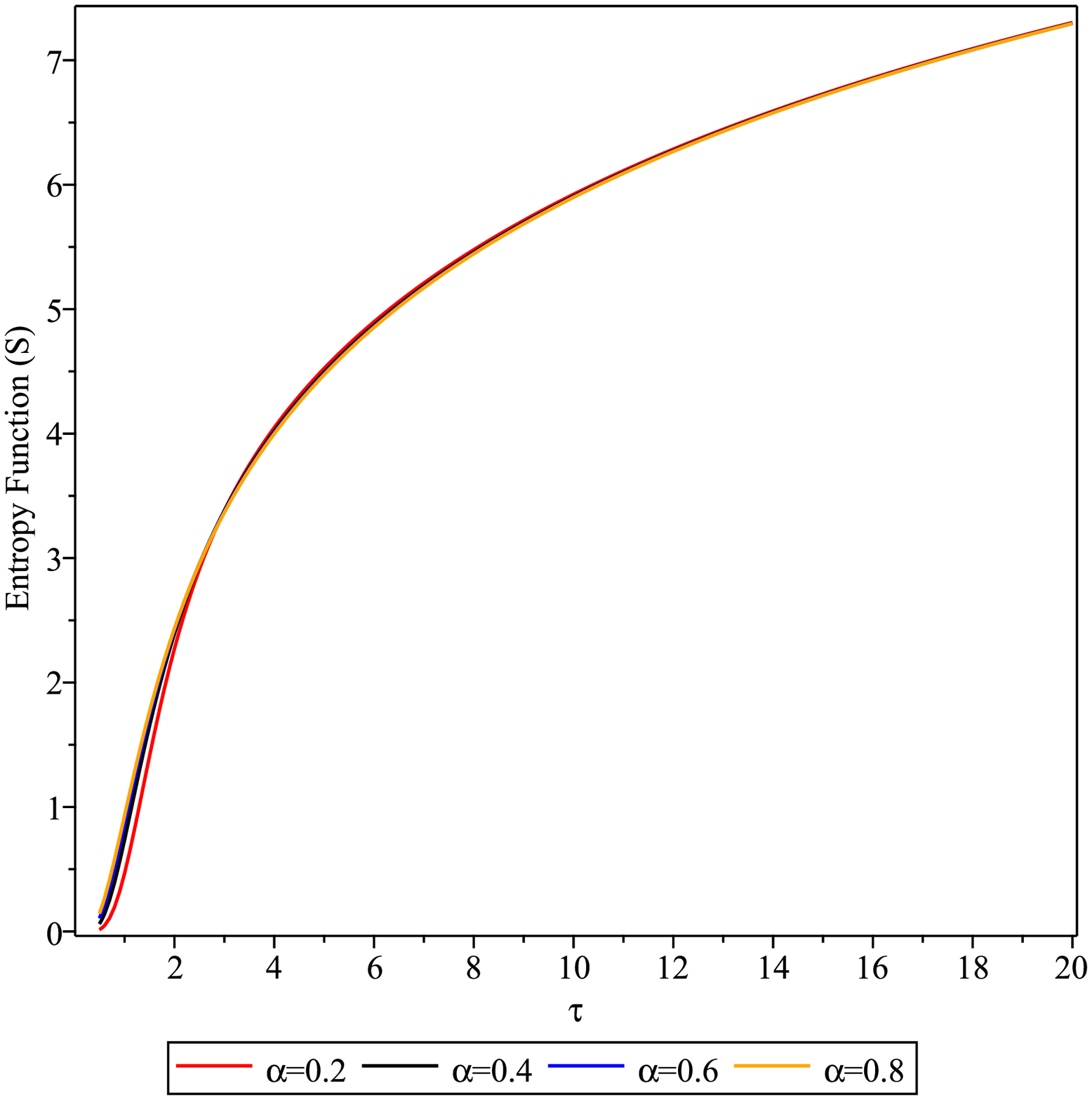}}\subfloat[Specific heat $C_{V}$ versus the reduced temperature $\tau$]{\centering{}\includegraphics[scale=0.3]{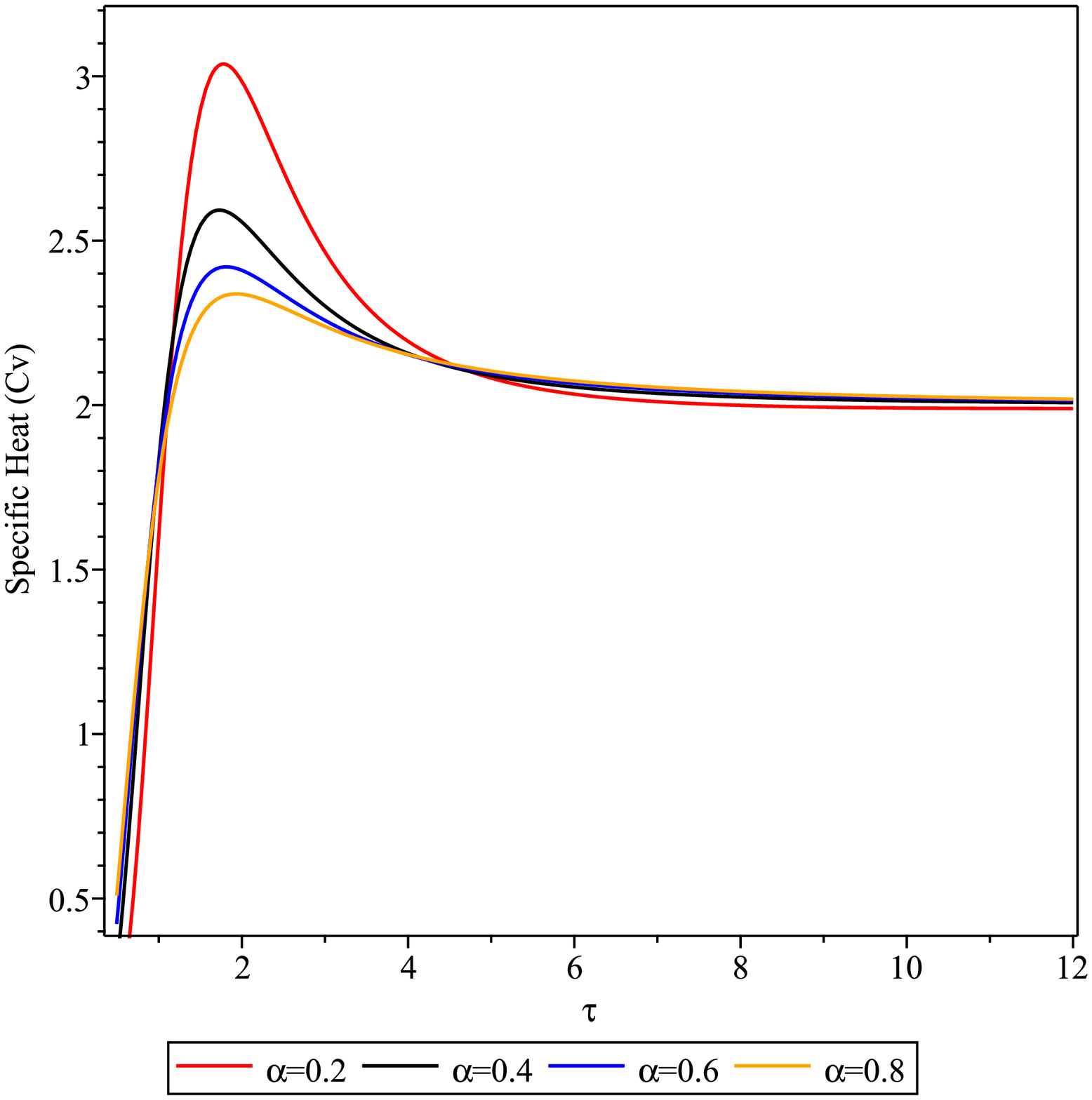}}
\par\end{centering}
\caption{\protect\label{fig:5}Statistical functions of KGO in (2+1)-dimensional
rotating cosmic string versus $\tilde{\beta}$ for $a=0.1$ and fixed
parameters $j=m=\omega=1.$}
\end{figure}

Fig. \eqref{fig:5} shows the different thermal quantities of KGO
in (2+1)-dimensional rotating cosmic string for $a=0.1$ derived from
the partition function keeping $n_{max}=1000$ terms of the series
in Eq. (\eqref{eq:88-1}). Following the figure, all thermal quantities
are plotted versus a reduced temperature $\tau$: here from the curves
of the numerical entropy function, no abrupt change, around $\tau_{0}$
has been identified in the curves of specific heat. This means that
the curvature, observed in the specific heat curve does not exhibit
or indicate an existence of a phase transition around a $\tau_{0}$
temperature. In addition, the effect of the parameter $\alpha$ for
fixed $a$ is well observed in the curves of specific heat. These
curves increased when $\alpha$ decreases.

\begin{figure}[H]
\begin{centering}
\subfloat[Free energy $F$ versus the reduced temperature $\tau$]{\centering{}\includegraphics[scale=0.3]{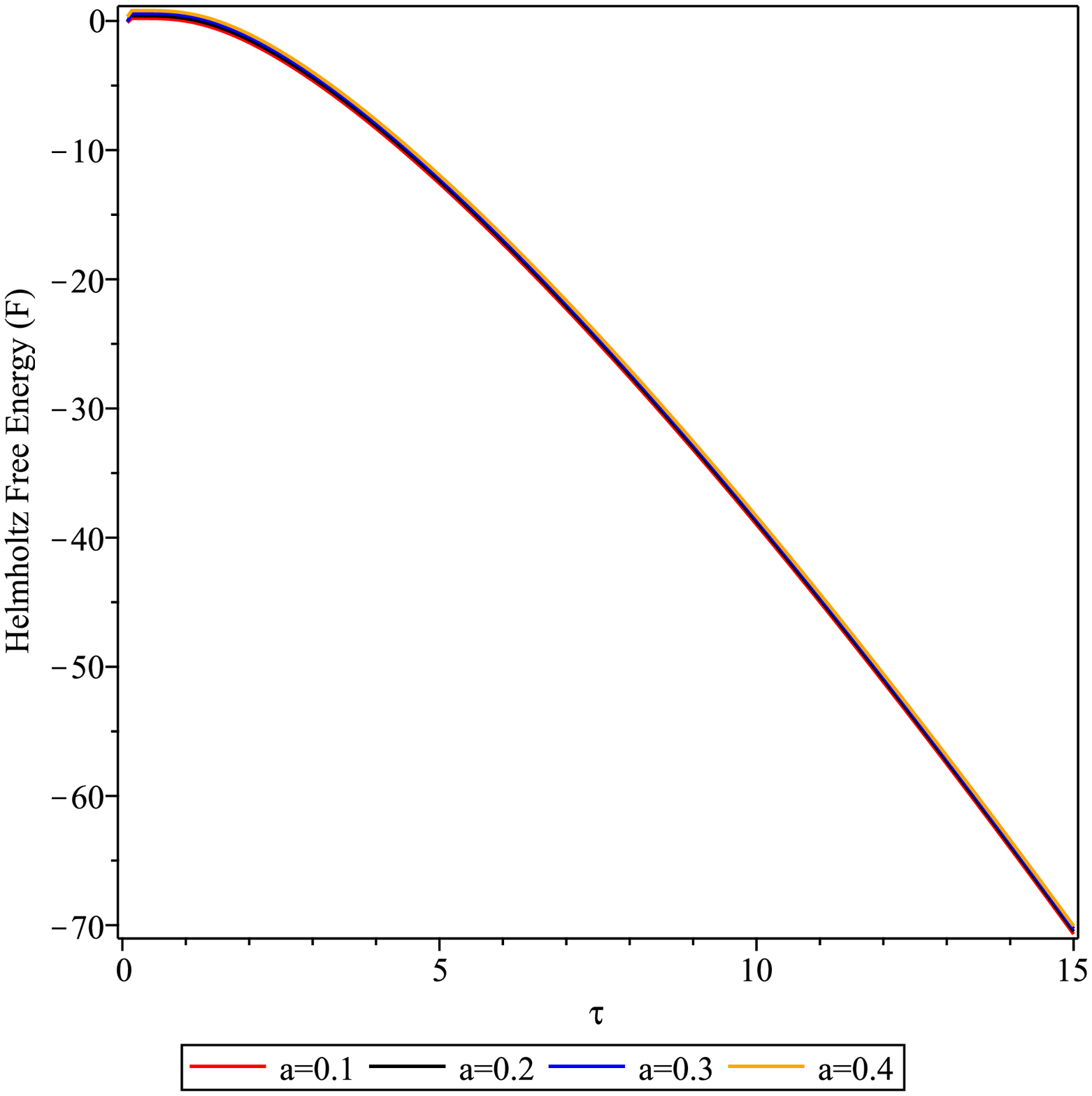}}\subfloat[Total energy $U$ versus the reduced temperature $\tau$]{\centering{}\includegraphics[scale=0.3]{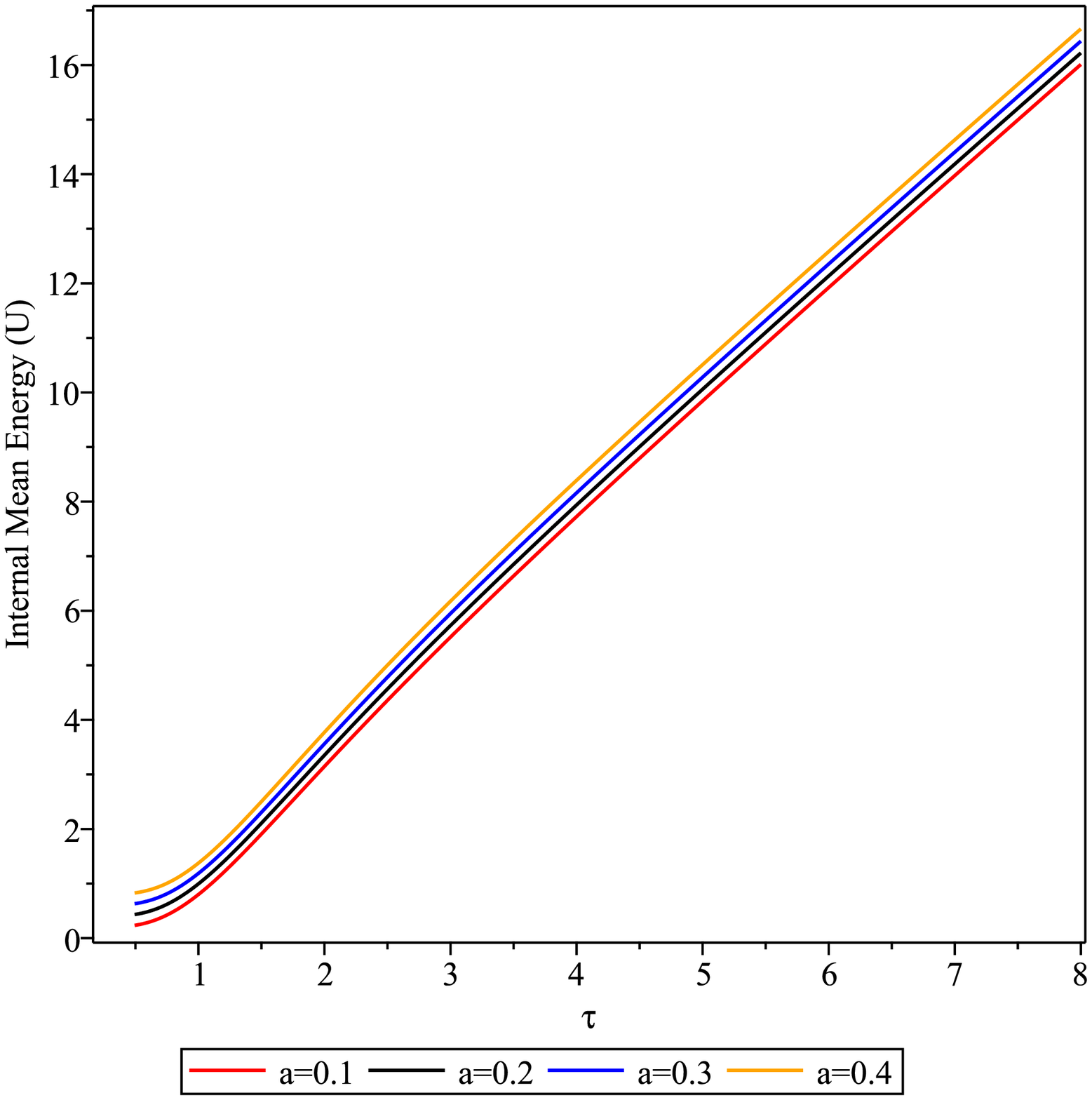}}
\par\end{centering}
\begin{centering}
\subfloat[Entropy $S$ versus the reduced temperature $\tau$]{\centering{}\includegraphics[scale=0.3]{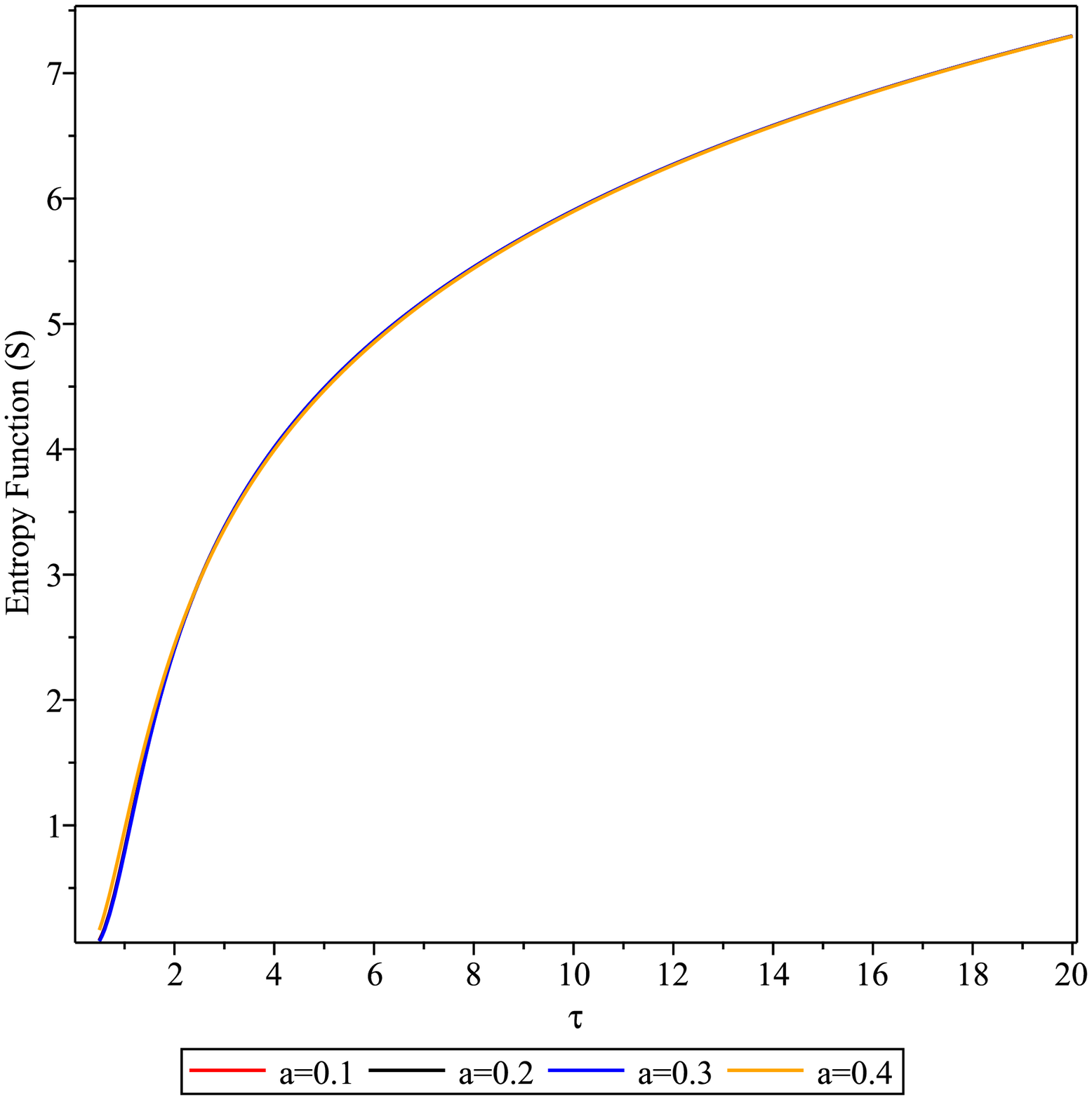}}\subfloat[Specific heat $C_{V}$ versus the reduced temperature $\tau$]{\centering{}\includegraphics[scale=0.3]{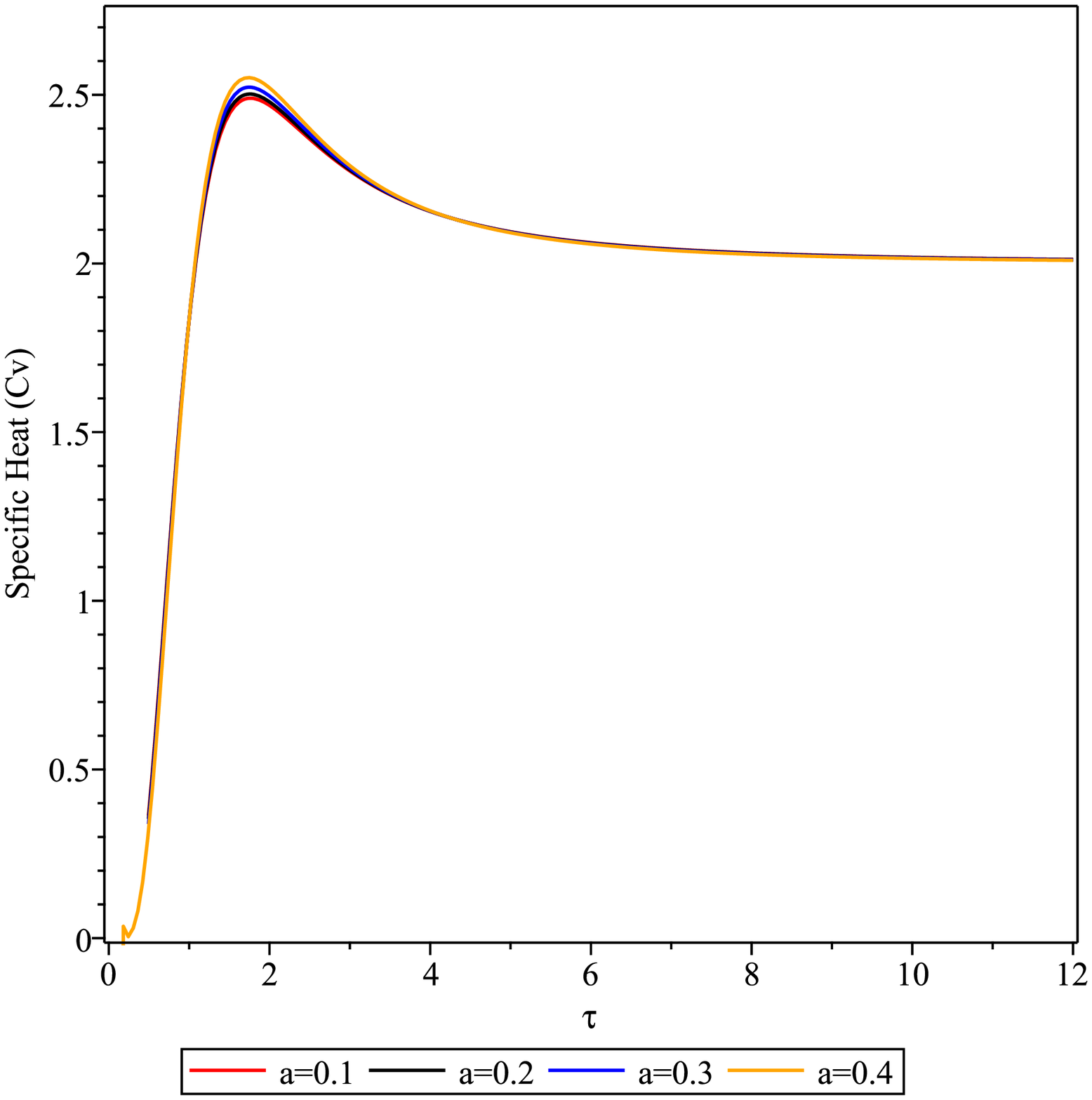}}
\par\end{centering}
\caption{\protect\label{fig:6}Statistical functions of KGO in (2+1)-dimensional
rotating cosmic string versus $\tilde{\beta}$ for $\alpha=0.5$ and
fixed parameters $j=m=\omega=1.$}
\end{figure}

Fig\eqref{fig:6} shows the same quantities but for fixed $\alpha=0.5$
and different values of $a.$ As we can see, all the curves of the
specific heat coincide a round of $\tau_{0}$ temperature. So, the
effect of the rotation on the thermal properties of our oscillator
has no effect.


\section{Conclusion }

The aim of the present work is to study the relativistic dynamics
of spinless quantum particle via the Feshbach-Villars representation
of two models, namely, the interaction of KGO with the gravitational
field generated by the background geometry of : a) static cosmic strings
and b) rotating cosmic strings. Starting from the revising the FV
formulation of scalar fields in Minkowski space-time, we derived the
corresponding formulations in two different curved manifolds. We obtained
the exact solutions of both systems and we presented the quantized
energy spectra which depend on the parameters that characterize the
space-time topology. It is not surprising to find that the wave-function
of our quantum system are expressed in terms of the confluent hyper-geometric
functions for both static and rotating cosmic strings, since the former
can be described throughout the previous one by choosing appropriate
coordinate transformation. In our above discussion, we made use of
the generalized Feshbach-Villars transformations in order to get relevant
observables. Such transformations have been shown to be exact and
covers any inertial and gravitational fields. In the last section,
we have examined the thermal properties of the system in question
by computing the partition function in terms of the Hurwitz zeta function.
We have noticed that these thermal quantities are affected by the
geometrical and topological parameters of the background geometry.

\section{Funding}
\begin{itemize}
\item Authors report that no funding is available.
\end{itemize}

\section{Author Contribution}
\begin{itemize}
\item Abdelmalek Buzenada concentrates on programming the system's thermal
properties.
\item Abdelmalek Boumali reviews, writes and completes the main text of
the manuscript
\item All authors reviewed the manuscript
\end{itemize}

\section{Data Availability Statemen}
\begin{itemize}
\item The datasets used and/or analysed during the current study available
from the corresponding author on reasonable request.
\end{itemize}

\end{document}